\documentclass[journal=inoraj,manuscript=article]{achemso}
\setkeys{acs}{articletitle = true}
\usepackage{chemformula}
\usepackage{xr}

\title{Chemical tuning of a honeycomb magnet through a critical point}
\author{Austin M. Ferrenti}
\affiliation{Department of Chemistry, The Johns Hopkins University, Baltimore, Maryland 21218, USA}
\alsoaffiliation{Institute for Quantum Matter, William H. Miller III Department of Physics and Astronomy, The Johns Hopkins University, Baltimore, Maryland 21218, USA}
\email{aferren2@jhu.edu}
\author{Maxime A. Siegler}
\affiliation{Department of Chemistry, The Johns Hopkins University, Baltimore, Maryland 21218, USA}
\author{Shreenanda Ghosh}
\affiliation{Institute for Quantum Matter, William H. Miller III Department of Physics and Astronomy, The Johns Hopkins University, Baltimore, Maryland 21218, USA}
\author{Xin Zhang}
\affiliation{Department of Chemical and Biomolecular Engineering, The Johns Hopkins University, Baltimore, Maryland 21218, USA}
\author{Nina Kintop}
\affiliation{Fraunhofer Research Institution for Materials Recycling and Resource Strategies IWKS, Aschaffenburger Straße 121, 64357 Hanau, Germany}
\author{Hector K. Vivanco}
\affiliation{Department of Chemistry, The Johns Hopkins University, Baltimore, Maryland 21218, USA}
\alsoaffiliation{Institute for Quantum Matter, William H. Miller III Department of Physics and Astronomy, The Johns Hopkins University, Baltimore, Maryland 21218, USA}
\author{Chris Lygouras}
\affiliation{Institute for Quantum Matter, William H. Miller III Department of Physics and Astronomy, The Johns Hopkins University, Baltimore, Maryland 21218, USA}
\author{Thomas Halloran}
\affiliation{Institute for Quantum Matter, William H. Miller III Department of Physics and Astronomy, The Johns Hopkins University, Baltimore, Maryland 21218, USA}
\author{Sebastian Klemenz}
\affiliation{Fraunhofer Research Institution for Materials Recycling and Resource Strategies IWKS, Aschaffenburger Straße 121, 64357 Hanau, Germany}
\author{Collin Broholm}
\affiliation{Institute for Quantum Matter, William H. Miller III Department of Physics and Astronomy, The Johns Hopkins University, Baltimore, Maryland 21218, USA}
\alsoaffiliation{NIST Center for Neutron Research, Gaithersburg, Maryland 20899, USA}
\author{Natalia Drichko}
\affiliation{Institute for Quantum Matter, William H. Miller III Department of Physics and Astronomy, The Johns Hopkins University, Baltimore, Maryland 21218, USA}
\author{Tyrel M. McQueen}
\affiliation{Department of Chemistry, The Johns Hopkins University, Baltimore, Maryland 21218, USA}
\alsoaffiliation{Institute for Quantum Matter, William H. Miller III Department of Physics and Astronomy, The Johns Hopkins University, Baltimore, Maryland 21218, USA}
\altaffiliation{Department of Materials Science and Engineering, The Johns Hopkins University, Baltimore, Maryland 21218, USA}
\email{mcqueen@jhu.edu}

\begin{document}
\begin{abstract}
  \ch{BaCo2(AsO4)2} (BCAO) has seen extensive study since its initial identification as a proximate Kitaev quantum spin liquid candidate. Thought to be described by the highly anisotropic XXZ-$J_1$-$J_3$ model, the ease with which magnetic order is suppressed in the system indicates proximity to a spin liquid phase. Upon chemical tuning via partial arsenic substitution with vanadium, we show an initial suppression of long-range incommensurate order in the BCAO system to \emph{T}~$\approx$~3.0~K, followed by increased spin freezing at higher substitution levels. Between these two regions, at around 10\% substitution, the system is shown to pass through a critical point where the competing $J_1$/$J_3$ exchange interactions become more balanced, producing a more complex magnetic ground state, likely stabilized by quantum fluctuations. This state shows how slight compositional change in magnetically-frustrated systems may be leveraged to tune ground state degeneracies and potentially realize a quantum spin liquid state. 
\end{abstract}

\section*{Introduction}
In the development of materials possessing improved mechanical, optoelectronic, and magnetic properties, small compositional changes can radically change the behaviors exhibited. To address long-standing scientific challenges - such as the stabilization of typically fragile, long-range spin entanglement in a real quantum spin liquid (QSL) - subtle modifications to promising, yet ultimately flawed candidates may be necessary.\cite{savary2017disorder, furukawa2015quantum} The chemical tuning of a magnetic system, particularly where there exists a high degree of spin frustration, can provide a broader understanding of the competing exchange interactions which determine the degeneracy of the magnetic ground state and often results in the discovery of exotic quantum phenomena.\cite{pasco2019tunable} While direct substitution on the magnetic sublattice can yield important information about the type and magnitude of exchange occurring in a material, the realization of a stable spin liquid state in a magnetically frustrated system requires a more thorough understanding of other structural features that help mediate magnetic exchange. This is especially relevant for phases mimicking the prototypical Kitaev spin liquid (KQSL), where S~=~1/2 moments on the vertices of a honeycomb lattice experience only highly anisotropic, Ising-type interactions.\cite{Kitaev2006,Jackeli2009} The bond-dependent nature of Kitaev exchange interactions makes such systems particularly sensitive to minor structural change.

One of the most intensely studied honeycomb spin liquid candidates, \ch{BaCo2(AsO4)2}, consists of d$^7$, octahedrally-coordinated Co$^{2+}$ ions, possessing a pseudospin-1/2 ground state doublet.\cite{zhong2020weak} The honeycomb lattice in \ch{BaCo2(AsO4)2}, or BCAO, consists of edge-sharing, lightly-distorted \ch{CoO6} octahedra, which are believed to account for the high degree of planar anisotropy observed in its magnetic behavior at low temperatures (Figure 1a-c).\cite{regnault2018polarized} BCAO orders antiferromagnetically at \emph{T}$_\text{N}$~=~5.4~K, with dominant in-plane ferromagnetic (FM) exchange, and has been shown to undergo two magnetic phase transitions, from an incommensurate to commensurate ordered state and then to a fully polarized state, upon the application of a $\mu_0\text{H}$~=~0.26~T and $\mu_0\text{H}$~=~0.53~T in-plane field, respectively.\cite{zhong2020weak} This suppression of the magnetic order with the application of only a relatively weak field was initially believed to indicate the presence of fairly weak non-Kitaev interactions in the material, relative to similar systems, which led to its identification as a proximate KQSL candidate. However, recent experimental studies point to a non-Kitaev, highly anisotropic and geometrically-frustrated ground state, better described by the XXZ-$J_1$-$J_3$ model, as opposed to the JK$\Gamma$ model associated with the KQSL.\cite{halloran2022geometrical} The frustration in BCAO then arises due not to Kitaev-type bond-dependent interactions, but rather due to competition between ferromagnetic superexchange ($J_1$) and third nearest-neighbor antiferromagnetic ($J_3$) exchange between Co$^{2+}$ ions in the honeycomb plane (Figure 1e).

While its promise as a quantum spin liquid candidate is ultimately dampened by the fact that it magnetically orders, the closeness of BCAO to the ideal octahedral Co$^{2+}$ geometry was initially believed to indicate that the material itself may be chemically tuned to exhibit more ideal QSL behavior. In order to do so without otherwise disrupting the honeycomb geometry of the cobalt layer, two potential avenues for improvement via chemical substitution exist. The substitution of some small percentage of either the interlayer polyanion group (\ch{AsO4^{3-}}) or alkaline earth atom (\ch{Ba^{2+}}) may alter the local \ch{Co^{2+}} coordination environment enough to produce more perpendicular Co-O-Co bond angles, and subsequently more perfect octahedral coordination, that would allow for maximal Kitaev and minimal Heisenberg exchange in the honeycomb plane. In particular, substitution of the interlayer polyanion group is expected to more strongly influence the third neighbor exchange interaction $J_3$, and thus also the degree of frustration in the XXZ-$J_1-J_3$ model.  

Of the divalent, reliably non-magnetic ions, partial substitution of the 12-fold coordinate \ch{Ba^{2+}} would likely only be possible with either \ch{Sr^{2+}} or \ch{Pb^{2+}}, both of which would represent a roughly 13\% reduction in ionic radius – an increase in size, at least with simpler ions, would not be possible. A pure Pb-analogue of BCAO is not known to exist, and the reported \ch{SrCo2(AsO4)2} phase does not contain Co on a honeycomb lattice. \cite{osterloh1994kenntnis} More flexibility exists for partial As-substitution, with tetrahedrally-coordinated \ch{V^{5+}}, \ch{P^{5+}}, \ch{Sb^{5+}}, and \ch{Nb^{5+}} all being potentially viable. Of these, only the P-analogue has been previously synthesized, and existing studies thereof have been hindered by the higher stability of two competing polymorphs.\cite{david2013puzzling} The maximum amount of substitution without the introduction of major structural deformations would likely occur for \ch{V^{5+}}, as its tetrahedral metal-oxygen bond lengths are roughly analogous to that of \ch{As^{5+}}, whereas \ch{Nb^{5+}} and \ch{Sb^{5+}} would represent a large increase, and \ch{P^{5+}} a large decrease, respectively.

Here we report the synthesis and structural and magnetic characterization of a series of V-substituted BCAO compositions (\ch{BaCo2(AsO4)_{2-2x}(VO4)_{2x}}), 0.025~$\leq$~x~$\leq$~0.70. At low substitution levels, 0.025~$\leq$~x~$\leq$~0.09, long-range incommensurate order in the system is gradually suppressed to \emph{T}~$\approx$~3.0~K, before disappearing down to at least \emph{T}~=~0.4~K in the intermediate substitution region. At higher substitution levels, 0.20~$\leq$~x~$\leq$~0.70, signs of spin freezing become apparent and the system begins to behave similar to the fully-substituted \ch{BaCo2(VO4)2} phase. The x~=~0.10 composition appears to represent a critical point between these two regions, displaying signatures of strong spin correlations, minor changes to the cobalt coordination environment, and a suppression of the incommensurate magnetic ground state, likely due to enhanced quantum fluctuations.

\section{Results}
\subsection*{pXRD}

\begin{figure}
  \includegraphics[width=1.0\textwidth]{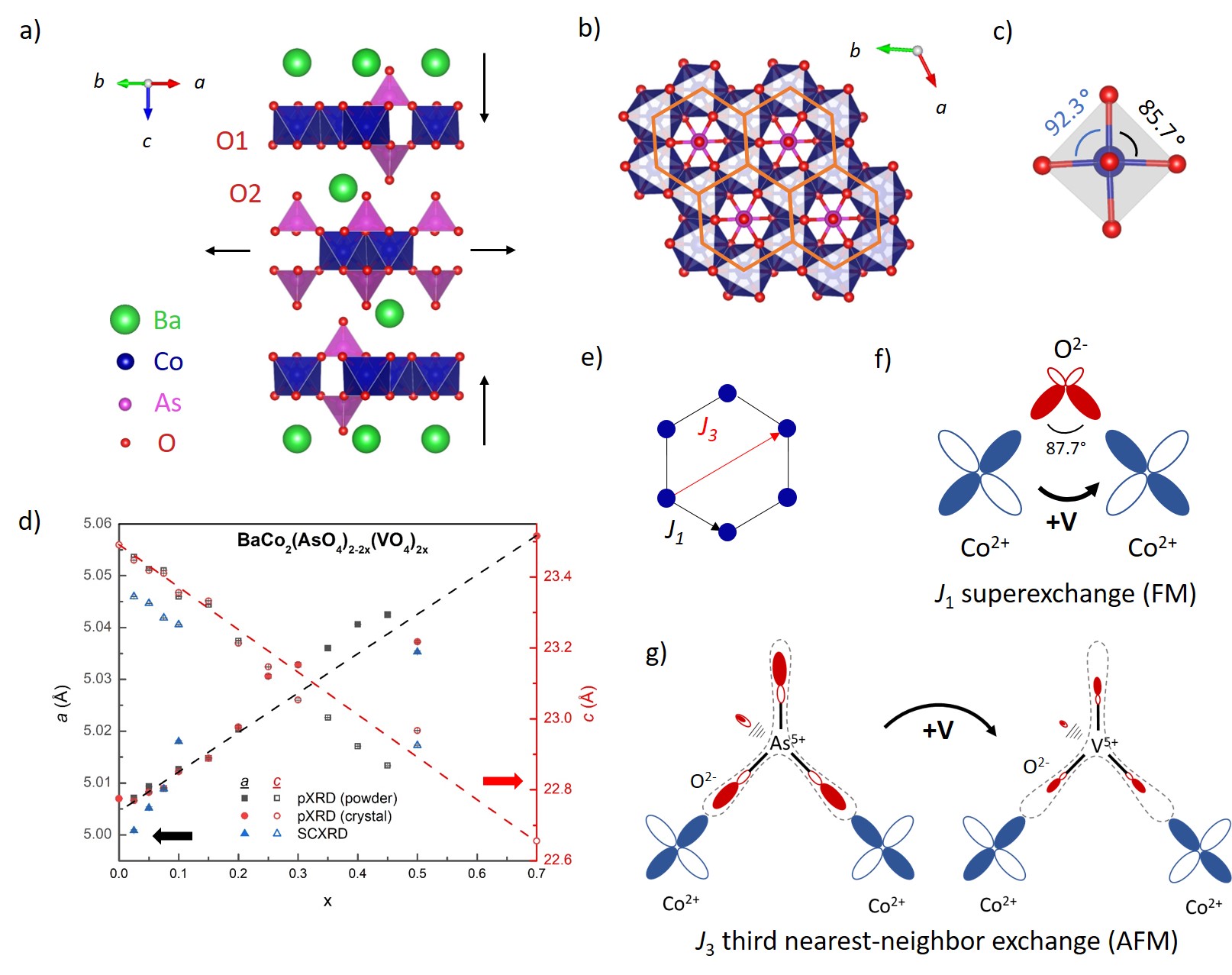}
  \caption{\textbf{Effects of vanadium substitution on the crystal structure of and exchange pathways in \ch{BaCo2(AsO4)2}.} The crystal structure of \ch{BaCo2(AsO4)2} (BCAO) projected a) within the \emph{ab} plane and b) along the stacking axis. c) Trigonal distortion of the cobalt octahedra within the honeycomb layer in pure BCAO. d) Calculated shifts in the \emph{a} and \emph{c} lattice parameters as a function of nominal V-substitution for As. Lattice parameters for each composition were derived from Rietveld refinement of once- (black) and twice-heated (red) pXRD data, as well as from refinement of twice-heated SCXRD data (blue). General trend lines are shown in black (\emph{a}) and red (\emph{c}). The calculated lattice parameter shifts for the nominally x~=~0.50 sample are lower than would be anticipated from the broader trends observed, indicating that more V was lost in the course of this synthesis. e) Dominant superexchange $J_1$ and third nearest-neighbor $J_3$ exchange pathways within the honeycomb plane described by the XXZ-$J_1$-$J_3$ model of geometric frustration. f) Orbital diagram of the $J_1$ superexchange pathway. On increasing V-substitution, this angle would be expected to increase and the magnitude of $J_1$ expected to decrease, relative to that of pure BCAO. g) Orbital diagram of the $J_3$ third nearest-neighbor exchange pathway. On increasing substitution of the strongly covalently-bonded arsenate group with the more ionically-bonded vanadate, exchange along this pathway is expected to weaken, reducing $J_3$.}
  \label{Struc}
\end{figure}

Initial attempts at partial arsenic substitution with vanadium in trigonal BCAO [space group \emph{R$\overline{3}$H} (148)] resulted in a series of compositions with clear, progressive shifts in both structural and magnetic properties. The powder diffraction patterns measured for each composition, shown in Figure S1, show shifts to higher 2$\theta$ for the majority of (h0l), (0kl), and (00l) reflections, and a slight shift to lower 2$\theta$ for the singular (hk0) reflection around 35.8 degrees. Assuming random incorporation of vanadium into the structure, Rietveld refinement of the experimental pXRD and SCXRD data in the \emph{R$\overline{3}$H} space group resulted in similarly good fits for all compositions upon initial and secondary heating (Tables S1-S4, S5-S10). Several compositions initially possessed a small ($\sim$3-5~wt\%) \ch{Ba3(AsO4)2} impurity phase, which was typically removed via a secondary heating step above the materials' melting point. Reflections corresponding to this impurity phase were also observed to shift with increasing nominal V-substitution of the target phase, indicating that partial substitution of arsenic with vanadium occurs for both phases present in the initial product mixture. Up to x~=~0.70, the roughly 0.02 Å increase in M-O bond lengths for tetrahedrally-coordinated vanadium is observed to produce a gradual contraction of the unit cell along the c-axis, as well as a gradual expansion of the a-axis, by 0.835 \AA~and 0.054 \AA~respectively (Figure \ref{Struc}d). Compared to BCAO, these relative shifts $\Delta c/c=-3.6\%$ and $\Delta a/a=1.1\%$ correspond to a volumetric contraction of $\Delta v/v=-1.4\%$. When substituted for the arsenate groups present in pure BCAO, the longer V-O bond lengths in the BCAO-V compositions produce both a compression of the interplanar region along the stacking axis and an additional tilting of the Co octahedra within the honeycomb plane. Beyond x~=~0.70, the tetragonal structure of the pure \ch{BaCo2(VO4)2} (BCVO) phase [space group \emph{I4$_1$/acd} (142)] begins to appear, with slight shifts in the observed reflections due to partial arsenic substitution for vanadium, and the trigonal BCAO structure gradually disappears. Similar to previous reports of vanadium substitution for phosphorus in \ch{BaCo2(PO4)2} (BCPO), the effective upper limit of vanadium uptake in the trigonal BCAO phase appears to be around 70\%.\cite{zhong2018field} 

More subtle structural changes on increasing vanadium substitution can be observed in the refined SCXRD data. Relative to pure BCAO, the average Co-O-Co bridging angle between neighboring \ch{Co^{2+}} ions increases by 0.7\%, while the average Co-O-As bridging angles decrease by -0.6\% and -0.2\%, respectively, up to x~=~0.50 (Figure S2). Laue diffraction patterns of representative BCAO-V single crystals, shown in Figure S3, were used to confirm the crystalline quality of each composition and show no signs of twinning or diffuse scattering, up to x~=~0.20.

The true degree of vanadium substitution in each composition was both estimated from refinement of the experimental diffraction data and confirmed via ICP-OES analysis. Calculated V-substitution levels obtained via Rietveld refinement of the As-site in BCAO after the first and second heating steps are shown in Figure S8a. Vanadium occupancies on the As-site typically refine close to nominal values for powder samples produced via initial heating of the reaction mixture, while the secondary heating step is observed to either meet or slightly exceed the expected occupancy. This can likely be attributed to the incorporation of the V-substituted \ch{Ba3(AsO4)2} impurity phase from the initial reaction mixture into the BCAO structure upon further annealing. 

ICP-OES analysis of synthesized BCAO-V samples agrees well with the V-substitution levels estimated via Rietveld powder and single crystal structural refinement, shown in Figure S8b and summarized in Table S12. No pronounced recovery of vanadium is observed from the initial powder synthesis to the secondary crystallization step, indicating that initial V-substituted \ch{Ba3(AsO4)2} impurity concentrations are actually generally minor. The measured V-content in the maximally-substituted composition suggests an upper limit of around 56.0\% before the tetragonal structure becomes more prevalent. A second sample of the x~=~0.10 composition was also analyzed and was found to possess closer to 17\% V. This sample was grown by the vertical Bridgman method, where the initial powder product was placed in a pointed alumina crucible, sealed under vacuum in a quartz tube, melted and slowly recrystallized. Due to the vertical temperature gradient, the slow re-solidification may have resulted in inhomogeneous distributions of vanadium throughout the grown crystal mass, and as a result these crystals were not used for further characterization.

As the shifts in both lattice parameters appear to be progressive over the studied composition range and the shift observed along the a-axis is small relative to that observed along the c-axis, it is likely that vanadium does not substitute for arsenic randomly in the structure, but rather as discrete domains. Attempts to generate accurate unit cells for each composition, particularly at low substitution levels, would thus be dependent on the manner in which these vanadium clusters are distributed throughout the structure. This would also imply a wider distribution of distinct cobalt coordination environments, producing a more complex magnetic ground state. Due to the difficulty in making this determination concretely, refinements on both pXRD and SCXRD data presented in this work were performed under the assumption of random substitution. Regardless of the true vanadium distribution, the progressive trends in the pXRD and SCXRD data and color changes in the grown crystals from pink (pure BCAO) to red (0~$<$~x~$\leq$~0.40) and gradually to black (x~$\geq$~0.40), together indicate average changes in the crystal field splitting energy of the octahedrally-coordinated Co and potentially also in the trigonal distortion thereof. 

\subsection*{Raman scattering spectroscopy}

Representative Raman scattering spectra collected from BCAO-V single crystals provide additional evidence for the gradual tuning of the Co coordination environment upon substitution of the interlayer polyanion group (Figure 2). In pure BCAO, the symmetry of the trigonal \emph{R$\overline{3}$H} structure is well-described by the $C_{3i}$ point group. Of the symmetrically-allowed optical modes, 18 are expected to be Raman-active (12 in the (ab) plane (6$A_g$ + 6$E_g$)) and 10 of which have been previously observed below 900 cm$^{-1}$.\cite{zhang2021and} The polarization dependence of band intensities was used to distinguish between $A_g$ and $E_g$ modes, which were observed to be strongest in the (x,x) and (x,y) configurations, respectively (Figure S4).\cite{zhang2021and} 

Upon partial substitution of As with V, this polarization dependence remains relatively consistent, with singlet $A_g$ modes initially predominating in the parallel polarization (x,x) configuration, doublet $E_g$ modes in the (x,y) configuration, and the overall measured intensity remaining higher for the former. The measured frequency shifts and linewidths for each vibrational mode as a function of substituted vanadium content are shown in Figure S5. Up to x~=~0.075, all $E_g$ and lower-energy $A_g$ modes begin to harden, while the $A_g$(5) mode begins to soften, all without notable broadening. The higher-energy $E_g$ modes also begin to decrease in intensity. 

At x~=~0.10, the intensity of the measured spectra in both configurations drops considerably, with that of the (x,y) polarization decreasing to less than half of that observed in either the x~=~0.075 or x~=~0.15 compositions. Interestingly, the intensity of the two highest-frequency $A_g$(5) and $E_g$(5) modes decreases by over an order of magnitude relative to the same samples, indicative of more considerable changes to the local bonding environment in BCAO. While reproducible across several single crystal samples, powder samples of this composition do not possess such a drop in magnitude, as shown in Figure S6. This effect likely then arises as a consequence of a distinctive clustering order for the x~=~0.10 phase that manifests only across larger domains. Beyond x~=~0.10, the majority of singlet modes begin to soften and the majority of the bands begin to markedly broaden. The higher-energy $E_g$ modes are not observed to shift in frequency, but rather decrease in intensity and are entirely absent by the x~=~0.20 composition. At and above this level of substitution, the intensity ratio of the remaining $A_g$ and $E_g$ modes is no longer dependent on the experimental configuration and the spectra differ only in magnitude. An additional broad feature also appears at $\sim$875 cm$^{-1}$ at higher substitution levels and is most prominent for the x~=~0.20 composition. This likely indicates a more heavily disordered bonding environment that introduces greater spin freezing in this region. 

Upon chemical substitution, the majority of systems display noticeable broadening of Raman active modes, indicative of random displacement of the substituted atom and subsequently higher disorder throughout the material.\cite{zhang2009doping, pena2020structural} While slight broadening of several modes is observed upon partial V-substitution of BCAO, the bands remain comparatively sharp, indicating that vanadium does not purely substitute randomly and further supporting that it may do so in a periodic manner.

Although not explicitly assigned in previous reports, the highest frequency of these likely correspond to breathing modes of the two distinct oxygen sites present in the structure.\cite{gohil2010raman} The frequencies at which these modes appear agree well with those previously attributed to honeycomb lattice materials containing edge-sharing Co-octahedra.\cite{husson1977normal,pena2020structural} As increasing vanadium incorporation is accompanied by a larger contraction of the unit cell along the stacking axis than expansion within the honeycomb plane, the oxygen vibrations occurring between cobalt layers would be expected to be more substantially impacted by greater substitution. The most impacted, highest-energy oxygen modes [$A_g$(5) and $E_g$(5)] are thus attributed to the apical oxygen of each arsenate/vanadate group not bound within the honeycomb plane, labeled O2 in Figure 1a. The lower frequency oxygen modes [$A_g$(4) and $E_g$(4)], then correspond to the oxygens directly bound to Co atoms, labeled O1 in Figure 1a. In the absence of broader structural change, these modes would be less impacted by substitution outside of the honeycomb planes.

The anomalous frequency and linewidth trends observed for vibrational modes in the x~=~0.10 composition, combined with the sharp decrease in measured intensity, provide evidence for a critical point at which signs of a more symmetric bonding environment emerge, despite the retention of the trigonal \emph{R$\overline{3}$H} structure suggested by the diffraction data. Whereas the gradual shifts observed in the calculated lattice parameters for each composition provide evidence for the amount of vanadium substituting into the BCAO structure, this critical point illustrates that the way in which that vanadium is incorporated is not consistent throughout the series. This composition clearly delineates the BCAO-V series between regions of order and disorder, driven by increased steric pressure about the honeycomb plane.

\subsection*{Magnetic characterization}

Temperature-dependent DC magnetic susceptibility measurements of V-substituted BCAO powder compositions, the ordering transition temperatures of which are shown in red in Figure 3, suggest a gradual tuning of the magnetic ground state of the system with increasing vanadium content. In pure BCAO, long-range incommensurate antiferromagnetic (AFM) order is suppressed above \emph{T}~=~5.4~K. From 0.025~$\leq$~x~$\leq$~0.09, the observed transition to the incommensurate ordered state decreases in temperature to \emph{T}~$\approx$~3.0~K, about half that of the parent material. From 0.10~$\leq$~x~$\leq$~0.20, no transition is typically observed, but rather the susceptibility rounds off and stagnates, down to at least \emph{T}~=~0.4~K. Above x~=~0.20, the transition reappears and gradually increases in temperature to exceed that of either pure BCAO or BCVO, before again decreasing to match that observed in the fully-substituted end-member with the dominance of the tetragonal-type phase.

To better understand how the magnetic ground state of the system evolves with increasing vanadium content, particularly for the low (x~=~0.025-0.20) V-substituted samples, temperature-dependent DC magnetic susceptibility measurements were also performed on single crystals of these compositions. With the field applied within the honeycomb plane ($\mu_0\text{H}\perp$ c, Figure S9), the incommensurate ordering transition declines to \emph{T}~=~4.06~K in the x~=~0.075 sample, slightly higher than that observed in the powder measurements. At and above x~=~0.10, the susceptibility instead plateaus below \emph{T}~=~3.0~K, indicating that the incommensurate ordered state is suppressed in these compositions. Some variability is observed for samples of the x~=~0.10 composition, with a weak antiferromagnetic transition sometimes appearing between \emph{T}~=~2.0~-~3.0~K before the susceptibility plateaus. As this occurred even for samples from the same batch and other measurement techniques did not show large variability, this can likely be attributed to the proximity of this composition to a critical point. The inherent instability of such a state leaves the magnetic behavior of these samples highly sensitive to a number of synthetic and handling conditions, such as furnace temperature gradient, cooling rate of the melt and sample handling, all of which can produce additional strains and alter magnetic ground state degeneracies. The higher-substituted samples also exhibit a bifurcation between the zero-field-cooled (ZFC) and field-cooled (FC) susceptibility measurements below \emph{T}~=~2.0~K, indicating some degree of spin freezing. The ZFC susceptibility for the x~=~0.15 composition trends negative below $T_\text{f}$, a feature typically attributed to materials containing multiple, highly anisotropic magnetic sublattices.\cite{kumar2015phenomenon} As \ch{Co^{2+}} is the only magnetic cation present in the BCAO-V system, this composition must possess a higher proportion of discrete cobalt clusters which produce a net negative magnetization at low temperatures.

Curie-Weiss analysis at high temperatures (\emph{T}~=~150~K~–~300~K) for $\mu_0\text{H}\perp$ c yields a Curie-Weiss temperature $\theta_{\text{CW}}$ $\approx$ 37(1)~K for all compositions and an effective magnetic moment $p_{\text{eff}}$ $\approx$ 5.5(2)~$\mu_B$ (Table 1, Figure S11), which lies between the effective moments 3.87~$\mu_B$ and 6.54~$\mu_B$ calculated for the $^4F_{9/2}$ term of the Co$^{2+}$ $3d^7$ ion in the spin only versus the full spin-orbital limits, respectively. The relatively stable positive Curie-Weiss temperature indicates the preservation of dominant FM coupling between adjacent \ch{Co^{2+}} ions. Yet $\theta_{\text{CW}}$ is $\approx$~20\% larger for the x~=~0.10 and x~=~0.15 samples, suggesting a net enhancement of ferromagnetism with V-substitution. The effective moment $p_{\text{eff}}$ for these compositions is also slightly lower than for either the other compositions or the parent BCAO, indicating additional minor quenching of the \ch{Co^{2+}} orbital angular momentum due to an increase in orbital symmetry. Although the substitution of the apical polyanion groups would not be expected to greatly impact the in-plane magnetic exchange pathways, these results, as well as the observed structural changes, indicate  modification of both the FM $J_1$ and the AFM $J_3$ exchange interactions and the degree of distortion of the cobalt octahedra.

When the field is instead applied along the stacking axis ($\mu_0\text{H}\parallel$ c), more substantial changes are observed in the measured DC magnetic susceptibility and calculated Curie-Weiss parameters (Figure S10). As in the in-plane measurements, increasing vanadium substitution results in a suppression of the incommensurate ordering transition to \emph{T}~=~4.06~K by x~=~0.075, beyond which the susceptibility plateaus below \emph{T}~=~3.0~K. Curie-Weiss analysis of the high-temperature susceptibility results in negative $\theta_{\text{CW}}$ of variable magnitude, as well as reduced effective magnetic moments, relative to pure BCAO (Table S13, Figure S11). All derived $\theta_{\text{CW}}$ for compositions in this range are smaller than that of the parent phase. The $\theta_{\text{CW}}$ for both the x~=~0.025 and x~=~0.20 crystals, \emph{T}~=~-67.8~K and \emph{T}~=~-62.8~K respectively, represent local minima in the series. Changes in superexchange pathways as well as in the spin-orbital crystal field levels could both be important factors. With increasing substitution, the effective moment $p_{\text{eff}}$ is more variable, however generally trends downward. This may arise due to the variable number of distinct cobalt coordination environments created as progressively more vanadium is incorporated into the structure. A wider distribution of coordination environments would possess more variability in the average degree of orbital quenching amongst \ch{Co^{2+}} ions in each honeycomb layer and potentially a lower net effective moment as a result.

A clearer comparison between the observed trends in relative magnetic interaction strength for spin components within and perpendicular to the honeycomb plane can be drawn from the Curie-Weiss-normalized temperature-dependent magnetic susceptibility of each composition, where \emph{C}/[($\chi-\chi_0$)~$|\theta|$]~=~\emph{T}/$|\theta|$~+~1 for dominant AFM interactions, and \emph{C}/[($\chi-\chi_0$)~$|\theta|$]~=~\emph{T}/$|\theta|$~-~1 for dominant FM interactions (Figure 4). For all compositions, spin components parallel to \emph{c} exhibit AFM deviations from Curie-Weiss behavior above the respective $T_{\text{N}}$, below which FM deviations are observed. These low-temperature deviations are attributed to the development of net AFM planar spin correlations, and decrease in magnitude with increasing vanadium substitution. When the field is applied within the honeycomb plane ($\mu_0H\perp c$), FM deviations from Curie-Weiss behavior dominate for temperatures above \emph{T}~$\approx$~150~K. Below this temperature, AFM correlations become more prevalent. The full suppression of the incommensurate AFM ordering transition for the x~=~0.10 composition is apparent in the absence of a thermal anomaly in both $\chi_{\parallel c}$ and $\chi_{\perp c}$.

A convenient measure of the magnetic anisotropy and its dependence on substitution and temperature lies in the susceptibility ratio, $\chi_{\perp}c/\chi_{\parallel}c$, which varies as a function of V-substitution (Figure S11c). In pure BCAO, $\chi_{\perp}c/\chi_{\parallel}c\approx$ 62, a value which drops considerably to 9 by x~=~0.025, before slowly increasing as a function of further substitution. At x~=~0.10, there is an additional slight decrease in the observed anisotropy. The trends in the magnetization measured along each of the principal axes are however not directly correlated. The magnetization gradually decreases when measured along the stacking axis with increasing V-content, while in-plane measurements are relatively consistent, with two sharper declines at x~=~0.025 and x~=~0.10. As no substitution occurs within the honeycomb lattice, but rather adjacent to it, little variation would be expected in the strength of the $J_1$ superexchange interactions between neighboring cobalt atoms in the \emph{ab} plane, though its spin-space anisotropy could change. With substitution of a minimal amount of vanadium into the structure and the corresponding contraction of the unit cell along the \emph{c} axis, Co$^{2+}$ ions within the honeycomb layer are pushed closer to and then further from perfect octahedral coordination. Further incorporation into and distortion of the parent structure progressively tunes the competing $J_1$/$J_3$ exchange interactions, with the x~=~0.10 composition representing a critical point at which the incommensurate state is suppressed without the introduction of considerable glassiness to the system. 

One feature commonly observed among magnetic cobaltates is the presence of metamagnetic transitions in field-dependent magnetization measurements, which are associated with the strong anisotropy of the Co$^{2+}$ Kramers doublet ground state.\cite{boonmak2011spin, leclercq2019metamagnetic, shizuya2007magnetic} In pure BCAO, this manifests as a pair of low-field metamagnetic phase transitions ($\mu_0\text{H}_{c1}$~=~0.26~T and $\mu_0\text{H}_{c2}$~=~0.53~T) which together produce the dumbbell-shaped hysteresis observed at \emph{T}~=~1.8~K.\cite{zhong2020weak} Field-dependent magnetization measurements performed on representative BCAO-V powder compositions, shown in Figure 5 as the derivative of magnetization with respect to the applied field strength, display a clear reduction in both the magnitude and applied field strength at which the \ch{Co^{2+}} metamagnetic transitions appear. Upon any amount of V-substitution, the dumbbell-shaped hysteresis observed in pure BCAO at \emph{T}~=~2.0~K is no longer present, with the two metamagnetic transitions gradually merging into one and broadening. At and above x~=~0.25, minor remanence reemerges and the system begins to behave more like the fully-substituted BCVO phase. The suppression of metamagnetism in this system is indicative of a gradual shift toward more perfect octahedral coordination and an increase in orbital symmetry of the Co$^{2+}$ cations. This would also be consistent with the suppression of the incommensurate ground state observed in pure BCAO. The disorder associated with V-substitution may also play a role, as it can impede phase transitions between the long wave length modulated spin structures associated with the metamagnetic transitions in pure BCAO. 

Measurements performed on lower V-content BCAO-V single crystal compositions provide additional insight into the evolution of the observed transitions with increasing substitution. In-plane ($\mu_0\text{H}\perp$ c) magnetization measurements (Figure 6) show no sign of hysteresis in any composition at either \emph{T}~=~2.0~K or \emph{T}~=~5.0~K, with all saturating by $\mu_0\text{H}$~=~$\pm$~1~T. At \emph{T}~=~0.4~K, only a weak remanence is present from x~=~0.10-0.20, which can likely be attributed to the weak FM coupling between neighboring Co$^{2+}$ ions within the honeycomb plane. The two sharp transitions observed in pure BCAO are broadened in all substituted compositions and gradually merge by the x~=~0.10 sample, leaving only a single smoothed curve. This indicates some alteration of the average cobalt coordination environment in the substituted structure upon increasing V-content and subsequently also to the ground state degeneracy of the system. Alternatively, the $J_1/J_3$ ratio may be passing through a critical value driven by V-substitution. Out-of-plane ($\mu_0\text{H}\parallel$ c) magnetization measurements show only a linear field-dependence and no notable hysteresis at both \emph{T}~=~2.0~K and \emph{T}~=~5.0~K, as in the parent material (Figure S12).

The evolution of the cobalt metamagnetic transitions in V-substituted BCAO is also observed in the in-plane, field-dependent AC magnetic susceptibility data, shown in Figure S13. At x~=~0.025, two transitions are visible at $\mu_0$H~=~0.25~T and $\mu_0$H~=~0.43~T, with the former being weaker. Upon further substitution, the higher-field transition gradually shifts to meet and merge with that occurring at lower field, in agreement with a gradual suppression of the incommensurate state. The magnitude of the combined transition then continuously decreases with increasing V-content. In contrast to the DC magnetization data, the magnitude of the AC susceptibility of several x~=~0.10 samples also notably decreases, relative to the other compositions, another indication of a more complex interplay between competing exchange interactions at low temperatures.

This discrepancy between the DC and AC response of several compositions is also apparent in temperature-dependent AC magnetic susceptibility measurements of BCAO-V powder samples (Figure S14). As in the DC susceptibility, substitution of vanadium for arsenic in BCAO results in a gradual shift in the observed transition temperature, down to \emph{T}~=~3.6~K in the x~=~0.075 composition, albeit with greater broadening. However, from x~=~0.10~-~0.20, a narrower transition appears from \emph{T}~=~2.5~-~2.7~K, lower in temperature than other members of the series. Above this level of V-substitution, the AC transition begins to weaken, broaden, and shift higher in temperature. When performed on single crystals of the less-substituted compositions, a similar trend appears, with variability again observed in the susceptibility of x~=~0.10 samples (Figure 7). Often samples of this composition possessed sharper transitions, relative to those of the x~=~0.075 and x~=~0.15 samples, providing additional evidence for a critical point in this compositional regime.

Notable differences between the magnetic response of a material via AC and DC characterization techniques typically indicate spin glass-like behavior, and subsequently varied spin relaxation rates in a material.\cite{snyder2001spin} In order to better understand the evolution of the magnetic ground state of the BCAO-V system, the temperature dependence of the AC magnetic susceptibility was measured as a function of applied AC frequency for the lower-substituted single crystal compositions (Figure 7, zoomed view in Figure S15). From x~=~0.025-0.075, there is no frequency dependence to \emph{T}$_\text{N}$ from $f$~=~50-10,000~Hz, whereas each of the x~=~0.10, x~=~0.15, and x~=~0.20 compositions display minimal shifts over the same range. The abrupt shift from the absence to appearance of a frequency-dependent \emph{T}$_\text{N}$ is a common signature of the onset of spin freezing, and when considered with the change observed in the DC susceptibility over the same range further suggests the presence of a critical point between the x~=~0.075 and x~=~0.15 compositions. Similar to the variability observed in DC susceptibility measurements, samples of the nominal x~=~0.10 composition display a variable frequency dependence of the AC susceptibility, both within and between batches. The typical frequency dependence of this composition is similar to that observed in the higher-substituted samples, however several crystals closer to the critical point display a reduced dependence. The comparative consistency observed across the other compositions further indicates a proximity to a critical point around x~=~0.10.

Plotting of this frequency dependence as 1/\emph{T} vs. ln($\omega$) yields a linear trend for each of the three higher V-content samples and indicates the presence of some degree of spin freezing (Figure S16, Table S14). While the uncertainty associated with determination of the precise transition temperature produces considerable error in the calculated activation energies and characteristic frequencies for the x~=~0.10, x~=~0.15, and x~=~0.20 compositions, both trend upward with increasing x. From x~=~0.10-0.20, linear fitting yields an Arrhenius-type activation energy between E$_\text{a}$/k$_B$~=~91.8~-~107.5~K and a characteristic frequency on the order of $\omega_0$~=~10$^{16}$~-~10$^{18}$~Hz, consistent with a cluster glass ground state. \cite{binder1986spin, kumar2020evidence, pakhira2020ferromagnetic} The magnetic ground state of the BCAO-V system is then likely best understood as a gradual suppression of the incommensurate order down to around x~=~0.10, followed by the introduction of increased glassiness after its disappearance, as has been commonly reported for phases possessing high degrees of magnetic frustration and substitution-induced disorder.\cite{kovnir2011spin} With respect to the interaction strength between magnetic Co$^{2+}$ ions in the BCAO-V structure, the x~=~0.10 composition then represents a critical point about which strong spin correlations persist down to low temperatures without the transition to the incommensurate state. The bifurcation of FC/ZFC magnetization data for \emph{T}~$\approx$~1.48~K (Figure S9d) indicates some degree of spin freezing at the critical point, as is generally expected in the presence of disorder. In the analogous \ch{BaCo2(PO4)_{2-x}(VO4)_x} system, similar behavior was attributed to the presence of a spin-liquid-like state stabilized by strong quantum fluctuations.\cite{zhong2018field}

Further evidence for a spin-glass-like ground state in this composition space is observed in the heat capacity of representative BCAO-V polycrystalline samples, shown in Figure 8. With increasing V-content, the transition attributed to the onset of long-range incommensurate order in pure BCAO again substantially broadens and declines in both magnitude and temperature, down to \emph{T}~$\approx$~4.0~K from x~=~0.075-0.20. In the absence of an applied field, all substituted compositions possess singular broad humps, higher in temperature than the transitions observed in both DC and AC susceptibility. Beyond this level of substitution, the transition shifts higher in temperature, in agreement with that observed in the AC susceptibility measurements and indicative of the reemergence of an ordered ground state in the system. 

Additional hints to the glassy nature of the magnetic ground state in the intermediate substitution regime can be observed in the field-dependent heat capacity of single-crystalline BCAO-V compositions. When the field is applied within the honeycomb plane ($\mu_0\text{H}\perp$ c), all compositions display a linear trend of further broadening and suppression of the transition to higher temperature, with the application of as small as a $\mu_0\text{H}$~=~0.5~T field being sufficient to partially suppress the magnetic entropy (Figure S17). This provides an additional indication of spin freezing in the magnetic ground state and is consistent with the observed stability of the in-plane FM correlations upon V-substitution. Conversely, application of the field along the stacking axis yields a more varied response, with progressively larger fields being required for any suppression of the feature to occur (Figure 9). In the x~=~0.10 sample, a $\mu_0\text{H}$~=~2.5 T field is sufficient to suppress the transition by about $\Delta T$~=~3.9~K, whereas for the x~=~0.15 sample this produces a shift of only $\Delta T$~=~1.7~K and for the x~=~0.20 sample this decreases to $\Delta T$~=~1.1~K. However when the applied field is increased to $\mu_0\text{H}$~=~5.0~T, the anomaly present for all three compositions is suppressed to a comparable degree. This suggests a finite barrier to spin reorientation which increases from the x~=~0.10 to the x~=~0.20 composition, supporting the presence of two distinct ground state phases - the latter behaving as a cluster glass and the former as a more complex intermediate between the regions of order and pure glassiness bounding the critical point.

Estimates of the magnetic contribution to the measured heat capacity were attempted via subtraction of the non-magnetic analogue of each V-substituted composition. Although \ch{BaMg2(AsO4)2} and its V-substituted derivatives could be easily synthesized, both direct subtraction and rational scaling of the Mg-analogues from the measured heat capacity of the cobalt phases ultimately did not suggest full recovery of the expected magnetic entropy by \emph{T}~=~50~K. Subsequent measurement of the Raman scattering spectrum of \ch{BaMg2(AsO4)2}, compared to that of BCAO in Figure S7, reveals that the substitution of the much smaller Mg atom for Co produces variable, non-uniform shifts in the frequencies of the respective vibrational modes. While the two phases share the same structure, the frequencies of phonon modes in pure and substituted BCAO then differ too considerably for reliable approximation via subtraction of the Mg-analogue. However, an estimate of the recovered magnetic entropy for each composition was obtained by scaling of the BMAO-V phase, such that the measured $C_{\mathrm{p}}/T$ converged with that of the Co-analogue by \emph{T}~=~50~K (Figure S18). Integration of this estimated contribution over the same temperature range yields an entropy rise greater than the predicted $\Delta S_{\mathrm{mag}}$~=~1/2Rln2 for an ideal Kitaev system and less than the $\Delta S_{\mathrm{mag}}$~=~Rln2 expected for a typical S~=~1/2 magnet.\cite{do2017majorana} Modeling of the evolution of the BCAO phonon spectrum as a function of V-substitution will ultimately be necessary to determine the true magnetic entropy rise in this temperature regime.

Reports of other substituted honeycomb phases exhibiting such intermediate spin relaxation rates have typically attributed this behavior to either the presence of stacking faults or magnetic site defects.\cite{wallace2015evolution} Due to the higher stability of the high-spin configuration in most Co$^{2+}$ compounds, as well as low likelihood of further oxidation to the Co$^{3+}$ oxidation state upon V-substitution, the presence of abundant magnetic site defects is unlikely. In either case, the calculated effective magnetic moment would also be expected to decrease more than is observed. For each composition, the substitution of clusters of arsenic atoms by vanadium would generate distinct stacking sequences within the structure akin to more random stacking faults, leading to more variability in the balance of $J_1$ superexchange and $J_3$ third nearest-neighbor interactions within the honeycomb plane. At low substitution levels, this increases the overall degree of geometric spin frustration, while at higher substitution levels this frustration is relieved by typical spin-freezing. Between these two regions, the stacking sequence in the x~=~0.10 composition produces a near-optimal tuning of the $J_1$/$J_3$ exchange interactions, fully suppressing the incommensurate-ordered state before the introduction of more considerable spin freezing. 

In solid solutions such as the BCAO-V system, qualitative changes in material properties are typically well described by percolation theory, wherein channels between impurity atoms form above some threshold impurity concentration.\cite{stauffer2018introduction} These channels produce deviations from the expected trends where impurity atoms enter the structure but largely do not interact with one another. On the two-dimensional honeycomb lattice, this threshold has been reported to be 0.7, which in the BCAO-V series is found to be the upper substitutional limit before transformation to the tetragonal \ch{BaCo2(VO4)2} structure.\cite{scher1970critical} However, magnetization and Raman scattering measurements show that the system passes through a critical point around x~=~0.10, suggesting that standard percolation theory can't be used to effectively describe the evolution of magnetic properties in this system. 

Several alternative models of exchange disorder via dilution or magnetic substitution have also been proposed for systems believed to possess dominant Kitaev exchange.\cite{cai2017magnetic,andrade2014magnetism} For the honeycomb iridate family of materials, these models typically agree well with experimental reports, where substitution tunes the magnetic ground state directly from long-range order to that of a typical spin glass. In $\alpha$-\ch{RuCl3}, substitution on the ruthenium site initially suppresses the long-range zigzag order before passing through a narrow region of short-range, quasistatic order and finally to the onset of increased spin freezing.\cite{lampen2017destabilization,do2018short,bastien2019spin} While substitution in these systems occurs on the magnetic honeycomb lattice, vanadium incorporation into the BCAO structure occurs only on the arsenic site bounding the honeycomb lattice and not for the cobalt within it. As our results are not well-described by any existing model for this type of substitution and direct honeycomb substitution in these related systems has also been claimed to tune the magnitude of off-diagonal exchange, the local minima observed in the x~=~0.10 composition likely arise from tuning of the $J_1$/$J_3$ exchange interactions via exchange disorder.

The impact of increasing disorder on the $J_1$/$J_3$ exchange can be understood as both a consequence of the gradual compression of the structure along the stacking axis and a reduction in the covalency of the polyanion orbitals mediating $J_3$ exchange. In compressing the structure along the c-axis, the Co-O-Co superexchange pathway responsible for the dominant FM first neighbor interaction is widened beyond the 87.7 degrees observed in pure BCAO (Figure S2a). As predicted by the Goodenough-Kanamori rules, superexchange through an oxygen bridge between half-filled orbitals is expected to be dominant FM when the bridging angle is 90° and gradually attain more AFM character as this angle nears 180°.\cite{goodenough1958interpretation,kanamori1959superexchange} As more vanadium is substituted into the structure, the average Co-O-Co bond angle widens, reducing the magnitude of the FM $J_1$ superexchange interaction. 

However, the weaker $J_3$ exchange proceeding through the more complex polyanion group can't be as easily rationalized by these rules (Figure 1g).\cite{halloran2022geometrical,das2021xy} In pure BCAO, the bounding arsenate group physically bridges third-nearest neighbor Co$^{2+}$ ions via a Co-O-As-O-Co pathway, however due to the large, delocalized orbitals of the arsenic center, bonding in the complex is highly covalent.\cite{bridgeman2001density} The polyanion group thus behaves akin to a superatom, with dominant superexchange-like interactions through the two oxygen atoms linking the cobalt and arsenic atoms. In BCAO, this interaction is AFM, and the competition between it and the FM $J_1$ exchange ultimately drives the transition to a long-range incommensurate state at low temperatures. In contrast to the strong covalency of the arsenate group, the smaller vanadate group is bonded more ionically, resulting in reduced orbital overlap and a more complex exchange pathway between third-nearest neighbor Co$^{2+}$ ions.\cite{bridgeman2001density} Increasing the vanadate content of the system would then also decrease the magnitude of the AFM $J_3$ exchange. This is supported by the decrease in the Co-O-As bond angles observed on increasing vanadium substitution (Figure S2b). While both $J_1$ and $J_3$ are expected to decrease with increasing vanadium substitution, the combination of the decrease in Co-O-As bond angle, the more ionic character of the introduced vanadate group, and the subsequently reduced orbital overlap suggests a faster decline in $J_3$. The progression of magnetic properties observed across the BCAO-V series thus indicates that the competition between the weakening $J_1$ and $J_3$ exchanges gradually tunes away the incommensurate ground state described by the XXZ-$J_1$-$J_3$ model, toward that of the standard easy-plane XXZ model with the more dominant FM $J_1$.\cite{das2021xy} The anomalous behavior observed about the critical point is then a consequence of a particular concentration and arrangement of vanadate groups within the structure where $J_1/J_3$ approaches a critical value and the incommensurate state is suppressed.

\section*{Discussion}
In conclusion, we present the systematic study of the structural and magnetic properties of honeycomb magnet \ch{BaCo2(AsO4)_{2-2x}(VO4)_{2x}} as it is tuned through a critical point via partial arsenic substitution with vanadium. At low substitution levels (0.025~$\leq~\text{x}~\leq$~0.09), magnetic susceptibility measurements show a gradual suppression of long-range incommensurate order in the system, down to \emph{T}~$\approx$~3.0~K. Around x~=~0.10, the incommensurate order disappears, minor spin freezing is introduced, and both the magnitude of the measured magnetic susceptibility and intensity of the measured Raman bands decrease considerably. Above this level of substitution, greater spin freezing is introduced, long-range order reappears, and the system gradually begins to behave akin to the fully-substituted \ch{BaCo2(VO4)2}. With increasing substitution of slightly-larger vanadium into the structure, the unit cell shrinks along the c-axis and the relative strengths of $J_1$ superexchange and $J_3$ third nearest-neighbor exchange can both be expected to decrease. Around the x~=~0.10 composition, $J_1$/$J_3$ approaches a critical value, producing a local minimum between the regions of long-range, incommensurate order and pure glassiness observed at lower and higher substitution levels, respectively. Likely stabilized by quantum fluctuations, this state represents a compelling example of how magnetically-frustrated systems may be chemically tuned to exhibit exotic quantum phenomena and ultimately realize a true quantum spin liquid state.

\section*{Materials and Methods}

Polycrystalline samples of V-substituted \ch{BaCo2(AsO4)2} were produced by the heating of well-ground mixtures of 2.9 mol \ch{BaCO3} (Strem Chemicals, 99.9\%), 2.0 mol \ch{Co3O4} (NOAH Technologies, 99.5\%, 325 mesh), and a combined 6.0 mol \ch{NH4H2AsO4} (Alfa Aesar, 98\%)/\ch{NH4VO3} (Beantown Chemical, ACS, 99.0\%). The mixtures were placed in an uncovered alumina crucible in air, heated to 305°C, held for 12 hours, heated to 875°C, held for 48 hours, cooled to room temperature, and finally heated to 925°C for 12 hours before again cooling to room temperature. All ramp rates were set at 100°C per hour. Both the 875°C and 925°C annealing steps, with an intermediate cooling to room temperature, were found to be necessary to minimize impurity phase formation. In each trial, the resulting product consisted largely of the expected \ch{BaCo2(AsO4)_{2-2x}(VO4)_{2x}} phase, as well as a small diamagnetic \ch{Ba3(AsO4)2} impurity. The omission of either the initial 875°C anneal or the intermediate cooling step resulted in a higher \ch{Ba3(AsO4)2} concentration and an additional \ch{Co3(AsO4)2} impurity. This also occurred when the stoichiometric 3.0 mol of \ch{BaCO3} was used in each synthesis, likely due to minor hydration of the \ch{Co3O4} precursor. Independent adjustments to the \ch{Co3O4} and \ch{NH4H2AsO4} concentrations were also attempted, but were not found to produce a purer product. As a result, representative \ch{BaCo2(AsO4)_{2-2x}(VO4)_{2x}} compositions were synthesized with a nominal 3.3\% \ch{BaCO3} deficiency.

Single crystals of each composition were produced via melting under static vacuum. Polycrystalline samples were added to an uncovered alumina crucible, placed in a 12mm ID, 16mm OD quartz tube, and sealed under vacuum. The tube was then heated at 1150°C for 12 hours, with heating and cooling rates of 100°C per hour. The resulting mass of crystals was then pried from the crucible and mechanically separated along the cleavage planes. Due to their layered morphology, the thinnest crystals were found to be the most reliably single-domain, however thicker crystals without noticeable signs of grain boundaries were also commonly produced. Powder and single crystal samples of the non-magnetic analogue of BCAO, \ch{BaMg2(AsO4)2} (BMAO) and its V-substituted compositions (BMAO-V) were produced by the heating and subsequent melting of 0.97 mol \ch{BaCO3} (Strem Chemicals, 99.9\%), 2.0 mol \ch{MgO} (NOAH Technologies, 99.99\%, 325 mesh), and a combined 2.0 mol \ch{NH4H2AsO4} (Alfa Aesar, 98\%)/\ch{NH4VO3} (Beantown Chemical, ACS, 99.0\%) under conditions identical to that described above for the BCAO-V series.

Similar substitution trials were also attempted with partial Nb, P, and Sb substitution for As, as well as with partial substitution of Pb and Sr for Ba. In all cases, only very minor shifts are observed in the measured powder X-ray diffraction patterns, owing to the lower probability of incorporation for ions much larger or smaller than those in pure BCAO. Initial magnetic characterization of several of these compositions is consistent with the larger distortions of the host lattice expected from partial substitution with the other species. The ordering transition is observed to occur at higher temperatures for all samples measured and for Nb-, Pb-, and Sb-substituted compositions a second, relatively broad transition appears at \emph{T}~=~2.0-3.0~K. This broad transition can likely be attributed to the freezing of spins in samples with more heavily-distorted Co-coordination environments.

Powder X-ray diffraction (pXRD) patterns were collected on a laboratory Bruker D8 Focus diffractometer with LynxEye detector and Cu K$\alpha$ radiation in the 2$\theta$ range from 5-120 degrees. All shifts in reflection position were initially determined relative to a silicon standard to ensure a reliable benchmark across compositions before subsequent Rietveld refinements. Refinements on pXRD data were performed using Topas 5.0 (Bruker). Calculated vanadium substitution levels were obtained via refinement of the arsenic site occupancy in each composition. The combined As and V occupancy on this site was constrained to one. All other site occupancies were also constrained to one. Structures were visualized with the Vesta 3 program.\cite{momma2011vesta}

Backscattered X-ray Laue diffraction was performed using a tungsten source operating at 10 kV and 10 mA with a Multiwire Laboratories MWL 120 real-time back reflection Laue camera.

Single crystal X-ray diffraction (SCXRD) measurements were performed using a SuperNova diffractometer (equipped with Atlas detector) with Mo K$\alpha$ radiation ($\lambda$~=~0.71073~\AA) under the program CrysAlisPro (Versions 1.171.41.93a and 1.171.42.49, Rigaku OD, 2020-2022). The same program was used to refine the cell dimensions and for data reduction. All reflection intensities were measured at \emph{T} = 213(2)~K. The structure was solved with the program SHELXS-2018/3 and was refined on \emph{F$^{2}$} with SHELXL-2018/3.\cite{sheldrick2015crystal} Analytical numeric absorption correction using a multifaceted crystal model (x~=~0.05, 0.075, 0.10, and 0.50 compositions) or empirical absorption correction using spherical harmonics (x~=~0.025 composition) were performed using CrysAlisPro. The temperature of the data collection was controlled using the Cryojet system (manufactured by Oxford Instruments). Calculated vanadium substitution levels were obtained via free variable refinement of the arsenic site occupancy in each composition. To account for the net decrease of 10 electrons on each vanadium-substituted arsenic site, the refined As-site occupancy was scaled by: 

\begin{equation} \label{eq:1}
x = (33-33z)/10
\end{equation} 

where $x$ is the estimated \% V-content for each sample and $z$ is the refined occupancy factor for As.\cite{sheldrick2015crystal} This accounts for the 10-electron difference between arsenic and vanadium, allowing for a more reliable estimate of V-occupancy on the As-site.

Optical emission spectroscopy based on inductively coupled plasma (ICP-OES) was used to experimentally determine the metal ratios in the samples. Crystals were dissolved in aqua regia at 240 °C (at 1200 W) for 1 h in a turboWAVE system from MLS-MWS. This process was repeated twice for each sample. The resulting concentration of Ba, Co, V, and As was measured 3 times and averaged using a PerkinElmer Optima 8300 ICP-OES spectrometer system. Additionally, the concentrations of Al, Ca, Mg, Na, and Si were recorded. 

Raman scattering spectra were measured in the backscattering geometry over the range from 100-1100 cm$^{-1}$ using a Horiba JY T64000 spectrometer equipped with an Olympus microscope, using a laser probe with a diameter of 2 microns and a laser power below 1 mW. The spectra were excited using the 514.5 nm line of a Coherent Innova 70C Spectrum laser. Representative spectra were collected with the $e_l$ polarization of the excitation light applied parallel to the \emph{ab}-plane of BCAO-V single crystals. Spectra were measured in parallel (x,x) and cross (x,y) polarizations, where x was oriented randomly in the (a,b) plane. No signs of laser-induced degradation of the samples were observed over the course of each measurement. All sample spectra were calibrated with a silicon standard to ensure a reliable comparison of frequency shifts on increasing vanadium substitution.

Magnetization data was collected on a Quantum Design Magnetic Property Measurement System (MPMS3). Magnetic susceptibility was approximated as magnetization divided by the applied magnetic field ($\chi\approx M/H$). AC magnetic susceptibility as a function of temperature ($\chi_{AC}$ vs. T) was measured for each powder composition with no applied DC field and an AC frequency of $f$~=~600~Hz. AC magnetic susceptibility as a function of applied DC field ($\chi_{AC}$ vs. B) was measured for each single crystal composition with an applied AC frequency of $f$~=~775~Hz. The temperature dependence of the AC magnetic susceptibility ($\chi_{AC}$ vs. T) for the single crystal compositions was measured on a Quantum Design Physical Property Measurement System (PPMS) with a $H$~=~20 Oe applied DC field, and an applied AC excitation field of $H$~=~1-10 Oe with frequencies of $f$~=~50~Hz, $f$~=~100~Hz, $f$~=~500~Hz, $f$~=~1000~Hz, $f$~=~5000~Hz, and $f$~=~10000~Hz. Assuming minimal interactions between neighboring clusters, the frequency dependence of the measured AC susceptibility was fit to the Arrhenius law, where E$_\text{a}$ is the activation energy for spin reorientation, k$_\text{B}$ is Boltzmann's constant, and \emph{T}$_\text{f}$ is the transition temperature observed:

\begin{equation} \label{eq:2}
\omega = \omega_0 \cdot exp(E_a/k_BT_f)
\end{equation} 

The linear fitting of a plot of 1/\emph{T}$_\text{f}$ vs ln($\omega$) for each composition yields E$_\text{a}$/k$_B$ as the slope and the characteristic frequency $\omega_\text{0}$ as the y-intercept.

Heat-capacity data was collected on a Quantum Design Physical Property Measurement System (PPMS) using the semi-adiabatic pulse technique with a 1\% temperature rise and over three time constants. The change in magnetic entropy as a function of temperature, $\Delta S_{\mathrm{mag}}$, was approximated as $\Delta S_{\mathrm{mag}}$~=~$\int C_{\mathrm{p}}/T\,dT$ of the measured $C_{\mathrm{p,mag}}$ for each composition ($C_{\mathrm{p, total, BCAO-V}}$ - $C_{\mathrm{p, total, BMAO-V}}$), from \emph{T}~=~0~-~50~K. The entropy rise from \emph{T}~=~0~-~2.0~K was estimated from linear extrapolation over this range. The measured $C_{\mathrm{p}}/T$ of pure BMAO and its 5\%, 10\%, 15\%, and 20\% V-substituted analogues was scaled by a factor of 1.20, 1.29, 1.44, 1.38, and 1.54, respectively, to ensure convergence with and subtraction from that of the Co-analogue by \emph{T}~=~50~K.

\newpage
\section*{Acknowledgments}
\textbf{Acknowledgements:} The authors would like to thank S. Bernier, B.W.Y. Redemann, and B. Wilfong for helpful discussions regarding analysis of Laue and powder diffraction, as well as magnetic susceptibility data. \textbf{Funding:}
This work was supported by the Institute for Quantum Matter, an Energy Frontier Research Center funded by the U.S. Department of Energy, Office of Science, Office of Basic Energy Sciences, under Grant DE-SC0019331. The MPMS3 system used for magnetic characterization was funded by the National Science Foundation, Division of Materials Research, Major Research Instrumentation Program, under Award \#1828490. TMM acknowledges support of the David and Lucile Packard Foundation. Chemical analysis was funded by the Fraunhofer Internal Programs under Grant No. 170-600006. \textbf{Author contributions:} A.M.F. synthesized and characterized the materials. A.M.F., S.G., X.Z., H.K.V., N.K., C.L., and T.H. performed the Raman spectroscopy, heat capacity, magnetic susceptibility, and ICP-OES experiments under the supervision of S.K., C.B., N.D., and T.M.M. A.M.F. and T.M.M. analyzed the data and interpreted the results. A.M.F. and T.M.M. wrote the manuscript, with contributions from all authors. \textbf{Competing interests:} The authors declare that they have no competing interests. \textbf{Data and materials availability:} All data needed to evaluate the conclusions in the paper are present in the paper and/or the Supplementary Materials. All data underlying this study will be made openly available at the online repository 10.34863/n3ds-9s54.

\newpage
\newpage

\begin{figure}
  \includegraphics[width=1.00\textwidth]
  {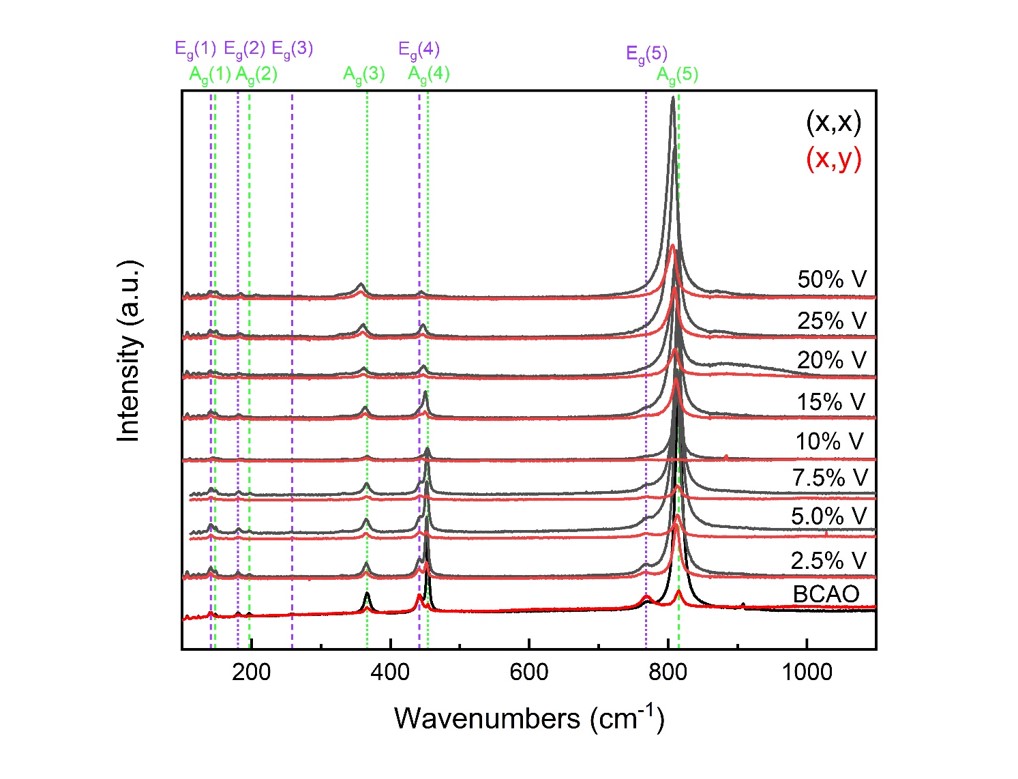}
  \caption{\textbf{Non-uniform shifts in vibrational modes on increasing vanadium substitution.} Measured Raman scattering spectra of pure and V-substituted BCAO with (x,x - black) and (x,y - red) polarized light. A$_g$ singlet modes are denoted by the green dashed lines, while E$_g$ doublet modes are denoted by the purple dashed lines. A$_g$ singlet modes are largely only active in the (x,x) polarization configuration, while E$_g$ doublet modes appear in both polarization configurations. All sample spectra were calibrated with a silicon standard to ensure a reliable comparison of frequency shifts on increasing vanadium substitution.}
  \label{Raman1}
\end{figure}
\newpage

\begin{figure}
  \includegraphics[width=1.00\textwidth]{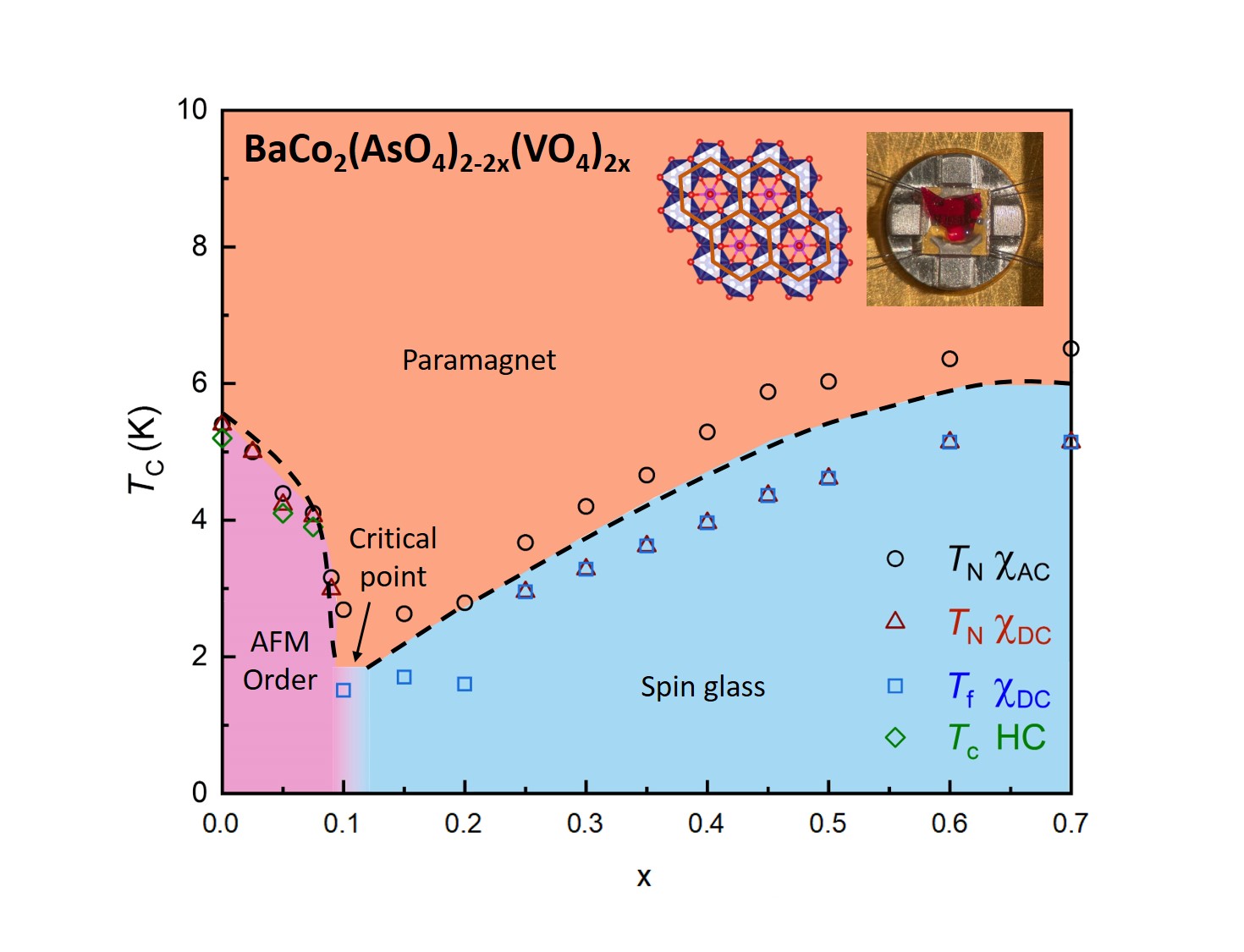}
  \caption{\textbf{Magnetic phase diagram of the \ch{BaCo2(AsO4)_{2-2x}(VO4)_{2x}} (0~$\leq$~x~$\leq$~0.70) series.} Derived from AC (\emph{T}$_\text{N}$~-~black circles) and DC (\emph{T}$_\text{N}$~-~red triangles; \emph{T}$_\text{f}$~-~blue squares) magnetic susceptibility, as well as heat capacity (\emph{T}$_\text{C}~-~$green diamonds) measurements of representative BCAO-V powder and in-plane ($\mu_0\text{H}\perp$ c) single crystal compositions. In the region from 0.025~$\leq~\text{x}~\leq$~0.09, all measurements suggest a gradual suppression of the incommensurate order observed in pure BCAO without the introduction of considerable glassiness. Above x~=~0.20-0.25, the system begins to behave gradually more like the parent \ch{BaCo2(VO4)2}, displaying an increase in all measured critical temperatures to above that observed in either parent material. In the intermediate range, 0.10~$\leq~\text{x}~\leq$~0.20, the system typically shows no sign of a transition to a long-range incommensurate state in the measured DC susceptibility, while still possessing weaker magnetic transitions in both AC susceptibility and heat capacity measurements. As the system is not known to undergo any structural phase transitions in this temperature regime, this likely indicates a suppression of the incommensurate ordered state and higher degree of magnetic frustration and ground state degeneracy for compositions in this range.}
  \label{Phase_diagram}
\end{figure}

\begin{table}
    \caption{\textbf{Fitting parameters obtained from analysis of BCAO-V single crystal magnetization measurements ($\mu_0\text{H}\perp$ c) from \emph{T}~=~150-300~K.}}
    \label{MvT_perp_CW}
    \begin{tabular}{c|c|c|c|c|c|c|c}
    \hline
    x & DC \emph{T}$_\text{N}$ & AC \emph{T}$_\text{N}$ & \emph{T}$_\text{f}$ & $\theta_{\text{CW}}$ & $p_{\text{eff}}$  & $\chi_0$ & $\chi_\perp$/$\chi_\parallel$\\ 
    & (K) & (K) & (K) & (K) & ($\mu_\text{B}$) & & \\\hline
    0 & 5.40 & 5.40 & - & 33.8 & 5.67 & - & 62 \\
    0.025 & 5.00 & 5.00 & - & 34.9(1) & 5.61(1) & -0.002 & 8 \\
    0.05 & 4.23 & 4.39 & - & 37.5(1) & 5.70(1) & -0.002 & 18 \\
    0.075 & 4.06 & 4.10 & - & 36.1(2) & 5.63(7) & -0.002 & 34 \\
    0.10 & - & 2.69 & 1.51 & 41.8(1) & 5.29(1) & 0 & 25 \\
    0.15 & - & 2.63 & 1.70 & 39.8(1) & 5.31(1) & 0 & 37 \\
    0.20 & - & 2.79 & 1.60 & 36.1(1) & 5.52(1) & -0.002 & 40 \\ \hline
    \end{tabular}
\end{table}

\begin{figure}
  \includegraphics[width=1.0\textwidth]
  {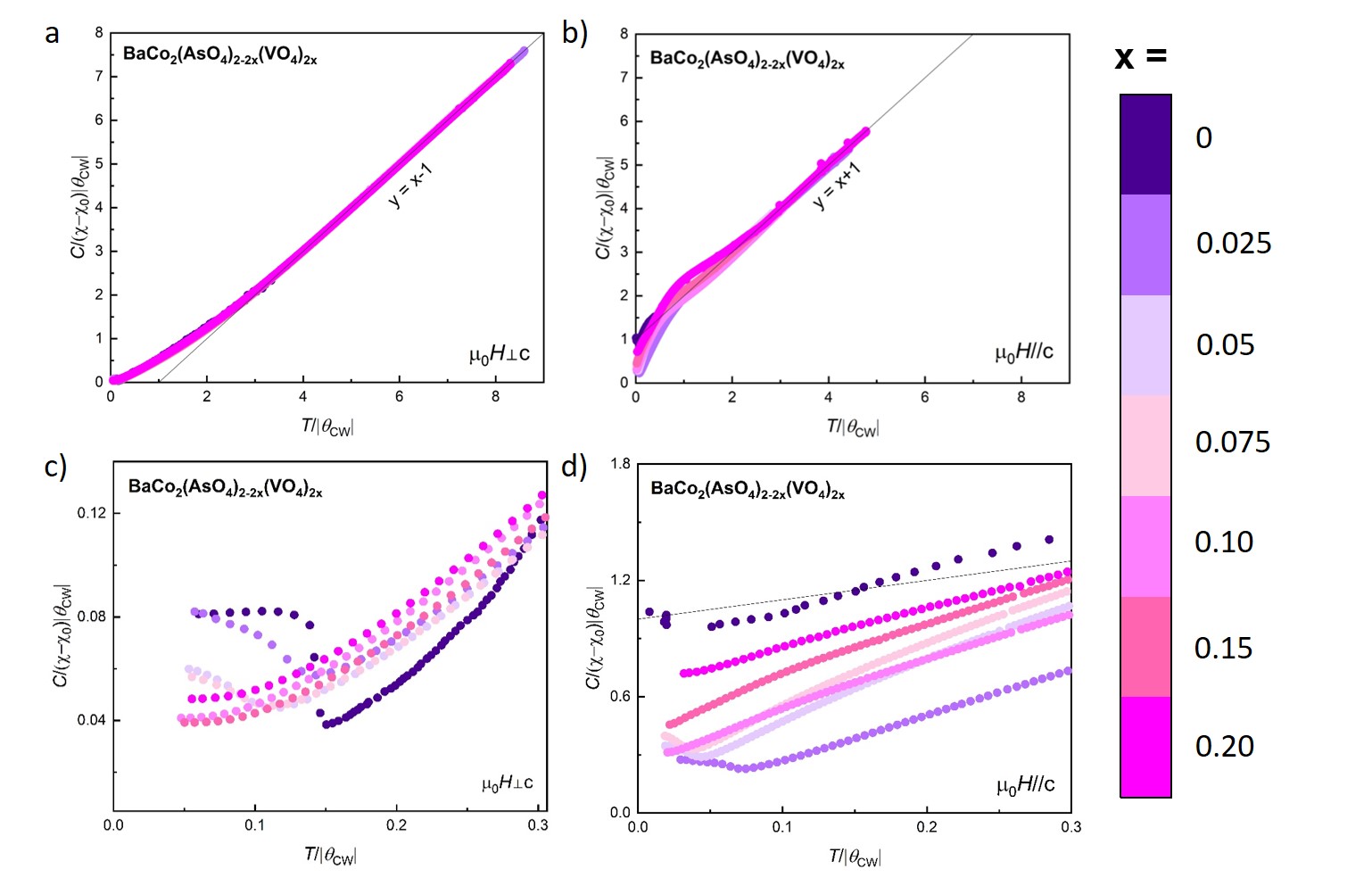}
  \caption{\textbf{Evolution of low-temperature deviations from Curie-Weiss behavior on increasing vanadium substitution.} Normalized inverse magnetic susceptibility of BCAO-V single crystals, where \emph{C}/[($\chi-\chi_0$)~$|\theta|$]~=~\emph{T}/$|\theta|$~+~1 for dominant antiferromagnetic (AFM) interactions, and \emph{C}/[($\chi-\chi_0$)~$|\theta|$]~=~\emph{T}/$|\theta|$~-~1 for dominant ferromagnetic (FM) interactions. Magnetic susceptibility data presented in previous works was extracted and scaled to fit the expected Curie-Weiss behavior at higher \emph{T} for unsubstituted \ch{BaCo2(AsO4)2}.\cite{zhong2020weak} When the field is applied a) within the honeycomb plane ($\mu_0H\perp c$), FM deviations from ideal behavior dominate in the c) low-\emph{T} regime. When the field is applied b) along the stacking axis ($\mu_0H\parallel c$), all compositions exhibit AFM deviations from ideal behavior just above their respective $T_{\text{N}}$, below which FM deviations are observed, which are attributed to the in-plane coupling between neighboring Co$^{2+}$ ions. With increasing V-substitution, these d) low-temperature FM deviations also decrease in magnitude. In both orientations, increasing V-substitution is observed to result in a full suppression of the AFM ordering transition by the x~=~0.10 composition.}
  \label{Norm_CW}
\end{figure}

\begin{figure}
  \includegraphics[width=1.0\textwidth]
  {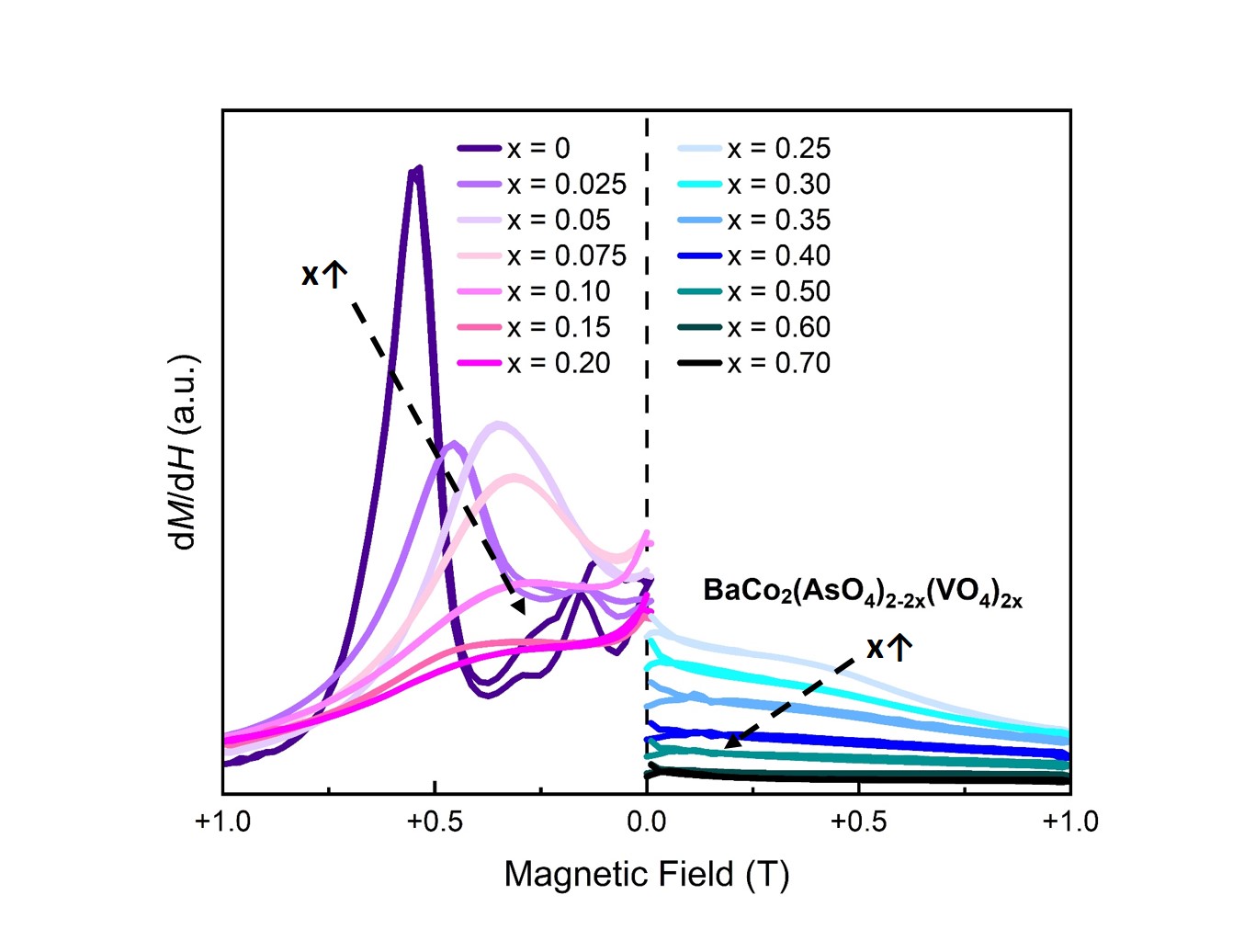}
  \caption{\textbf{Evolution of metamagnetism across the BCAO-V powder series.} Derivative of field-dependent magnetization with respect to field of representative x~=~0-0.70 BCAO-V powder samples at \emph{T}~=~2.0~K. For each composition, only one magnetization cycle is shown, however the behavior is consistent whether the samples are positively or negatively magnetized to $\pm$1~T. With increasing V-content, both the magnitude and field at which the cobalt metamagnetic transitions occur are observed to decrease, as shown by the black trend lines. No coercivity is observed from x~=~0.025-0.20. x~$\geq$~0.25 shown separately for clarity.}
  \label{dM/dH}
\end{figure}

\begin{figure}
  \includegraphics[width=1.00\textwidth]{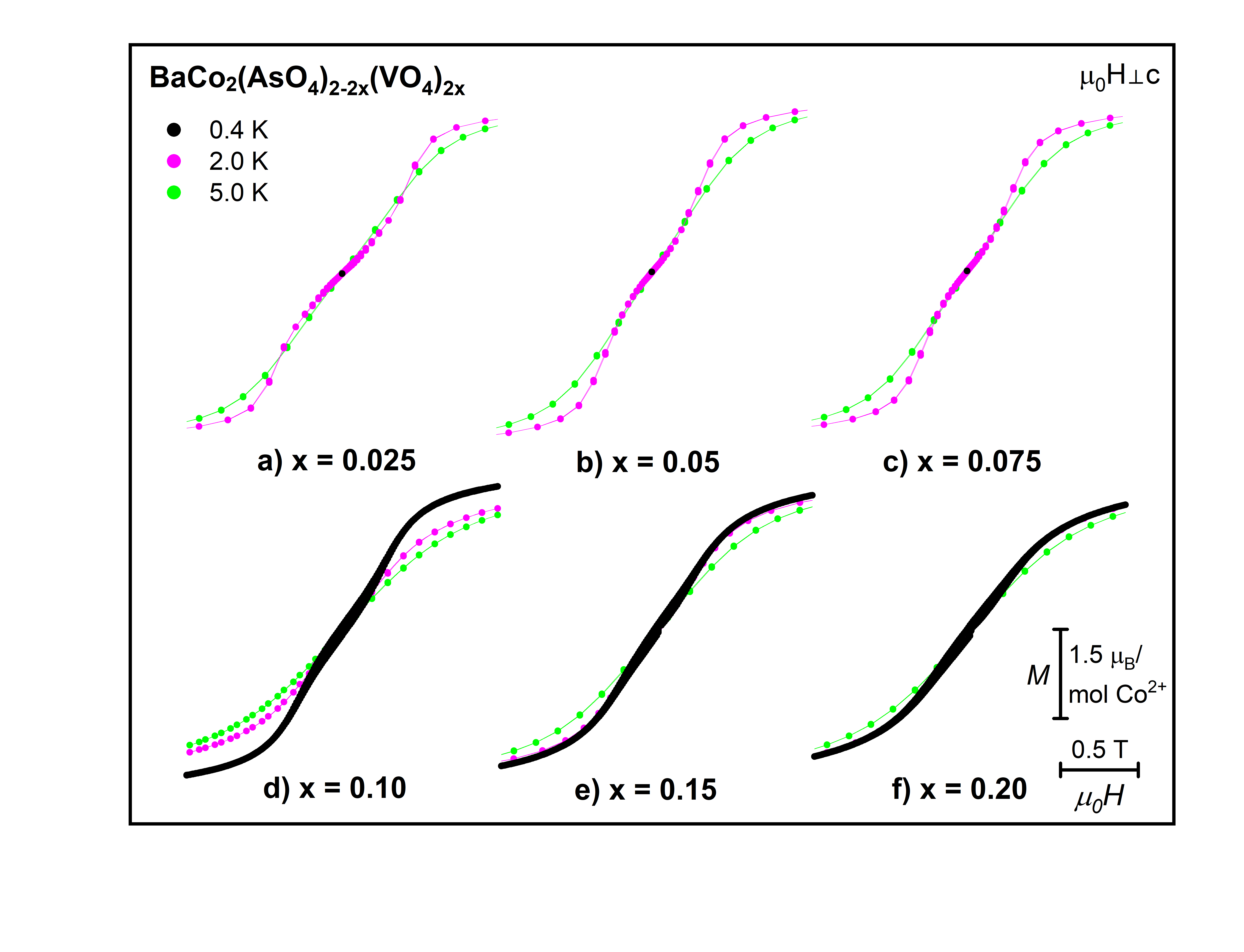}
  \caption{\textbf{Suppression of cobalt metamagnetic transitions in BCAO-V single crystals.} In-plane, field-dependent magnetization of representative a) x~=~0.025, b) x~=~0.05, c) x~=~0.075, d) x~=~0.10, e) x~=~0.15, f) x~=~0.20 BCAO-V single crystals at \emph{T}~=~0.4~K (black), \emph{T}~=~2.0~K (pink) and \emph{T}~=~5.0~K (green). Relative to the parent compound, no coercivity is observed for any composition at or above \emph{T}~=~2~K, and the cobalt metamagnetic transitions are gradually suppressed. Minor hysteresis is observed for the latter three compositions at \emph{T}~=~0.4~K, likely due to the in-plane ferromagnetic exchange  between neighboring Co$^{2+}$ ions. All compositions display a saturated magnetic moment by around $\mu_0\text{H}$~=~1~T, similar to the parent material.}
  \label{MvH_in_plane}
\end{figure}

\begin{figure}
  \includegraphics[width=1.0\textwidth]{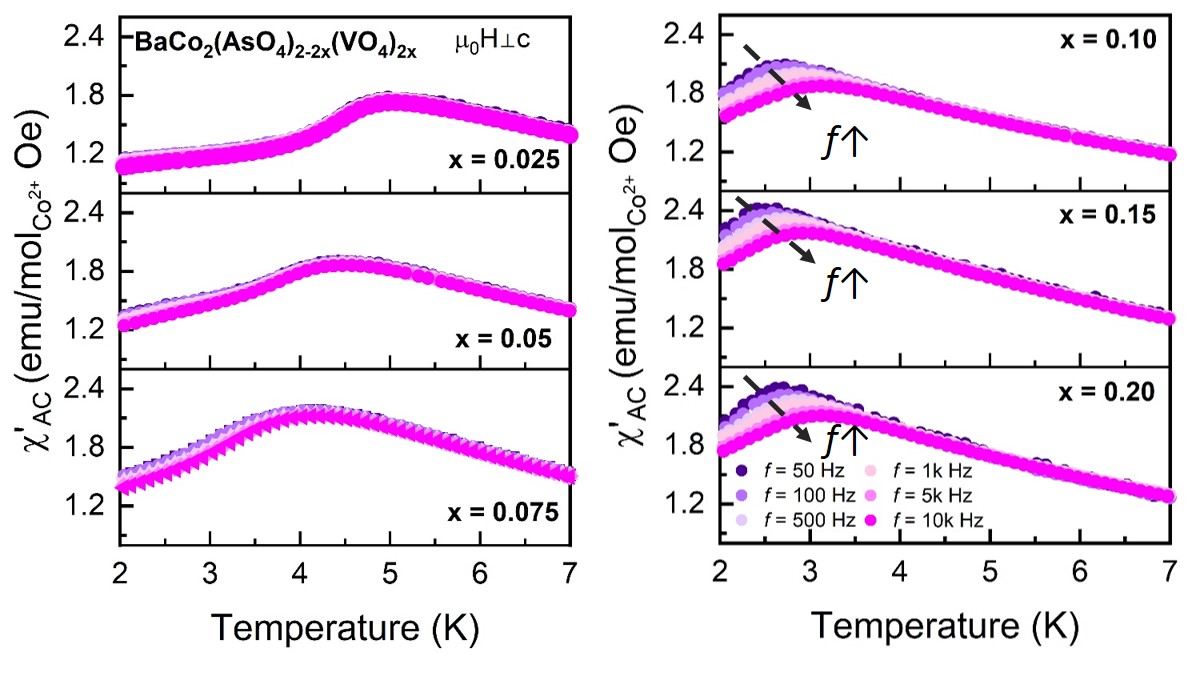}
  \caption{\textbf{AC frequency-dependence of BCAO-V single crystals on increasing substitution.} Real component of the AC magnetic susceptibility of (left - from top to bottom) x~=~0.025, x~=~0.05, x~=~0.075 and (right - from top to bottom) x~=~0.10, x~=~0.15, and x~=~0.20 BCAO-V single crystal compositions as a function of temperature from \emph{T}~=~2.0-7.0~K and with an applied AC field of frequency $f$ = 50-10000 Hz (purple to pink). From 0.025 $\leq x\leq$ 0.075, the transition temperature is observed to gradually decrease with increasing x, but is frequency-independent. For these compositions, an increasing frequency dependence is observed only below the transition temperature. Above this level of substitution, an increasing frequency dependence of the transition temperature is observed.}
  \label{AC_chi_vs_T_full}
\end{figure}
\newpage

\begin{figure}
  \includegraphics[width=1.00\textwidth]{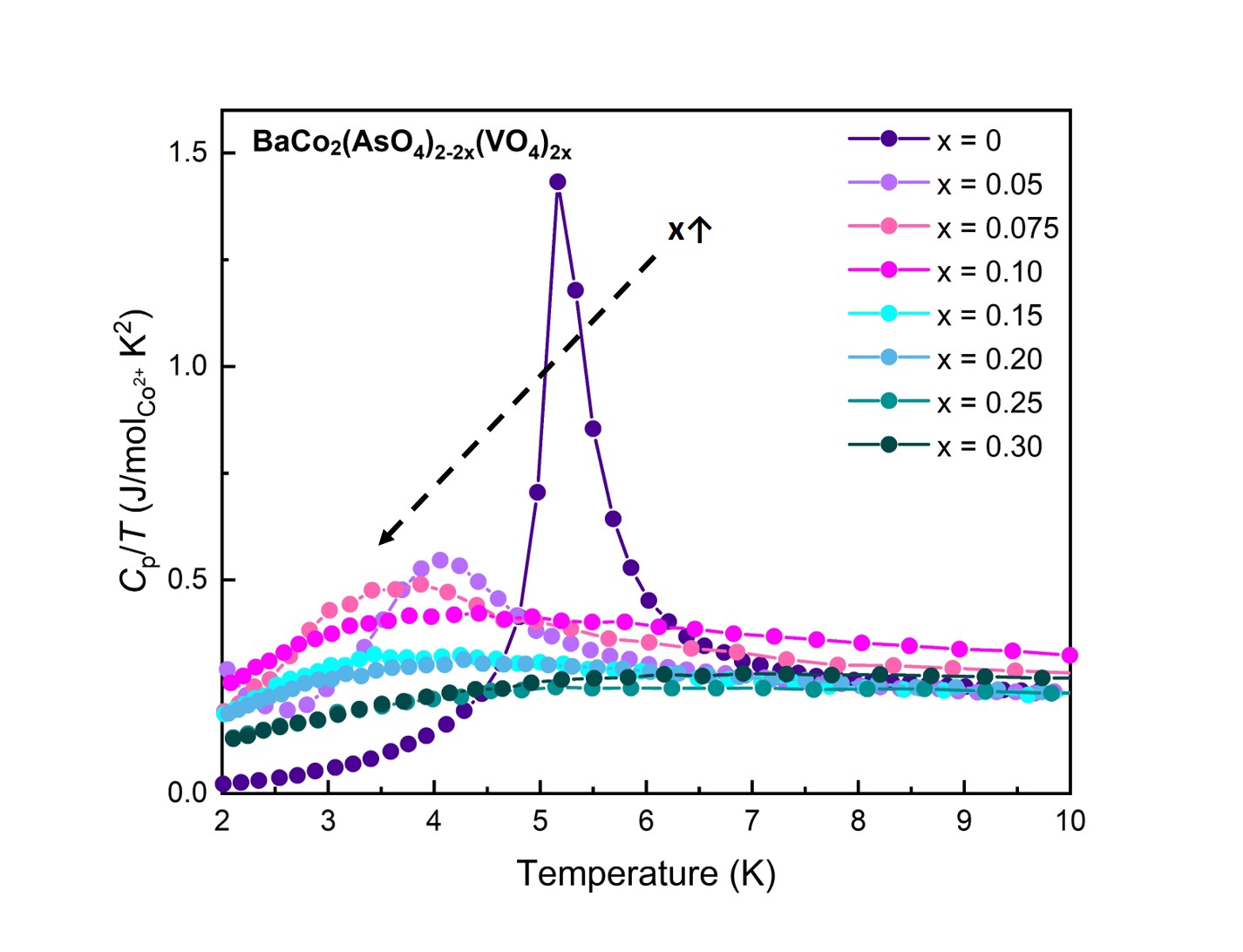}
  \caption{\textbf{Suppression of low-temperature heat capacity on increasing vanadium substitution.} Representative heat capacity of x~=~0-0.30 BCAO-V powder samples as a function of temperature from \emph{T}~=~2.0-10~K with no applied field. With increasing V-content, the magnetic ordering transition declines in both magnitude and temperature down to x~=~0.20 (as highlighted by the black dashed line). Above this level, the transition begins to shift back to higher temperatures, but remains broad and much weaker than in the parent material.}
  \label{HC_all_comb}
\end{figure}

\begin{figure}
  \includegraphics[width=1.00\textwidth]{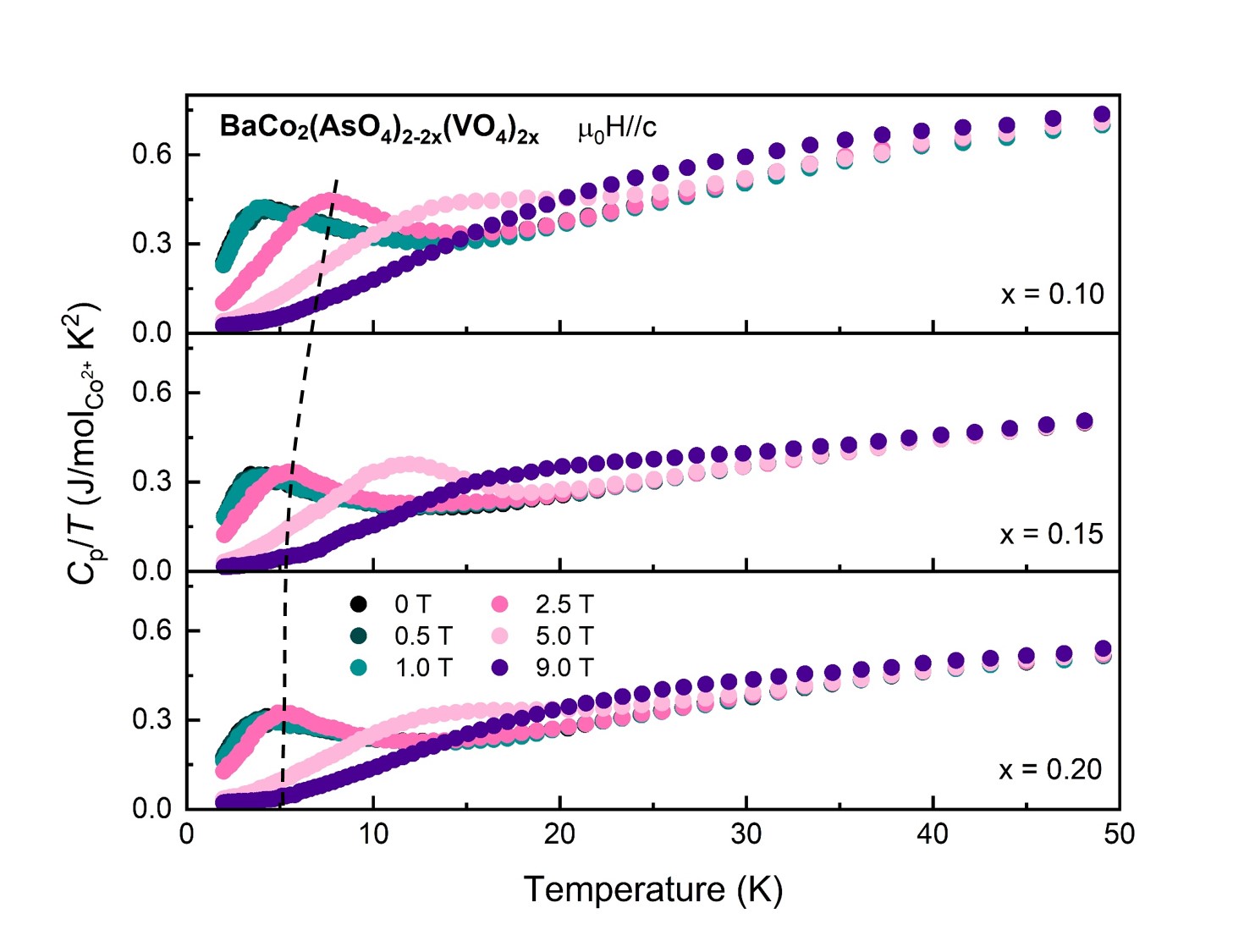}
  \caption{\textbf{Increasing barrier to spin reorientation in \ch{BaCo2(AsO4)_{2-2x}(VO4)_{2x}} (0.10~$\leq~\text{x}~\leq$~0.20).} Representative heat capacity of (from top to bottom) x~=~0.10, x~=~0.15, and x~=~0.20 BCAO-V single crystals as a function of temperature in a $\mu_0$H~=~0-9~T magnetic field applied along the stacking axis. Increasing V-content results in a higher barrier to suppression of the magnetic entropy in the system, as indicated by the shifting position of the feature observed for each composition when measured in an applied field of $\mu_0$H~=~2.5~T (highlighted by a dashed black line).}
  \label{HC_SC_para}
\end{figure}

\clearpage
\end{document}


\newpage
\tableofcontents
\newpage

\section{Structural characterization and refinement}
\subsection{pXRD refinement data}

\begin{figure}
  \includegraphics[width=0.9\textwidth]
  {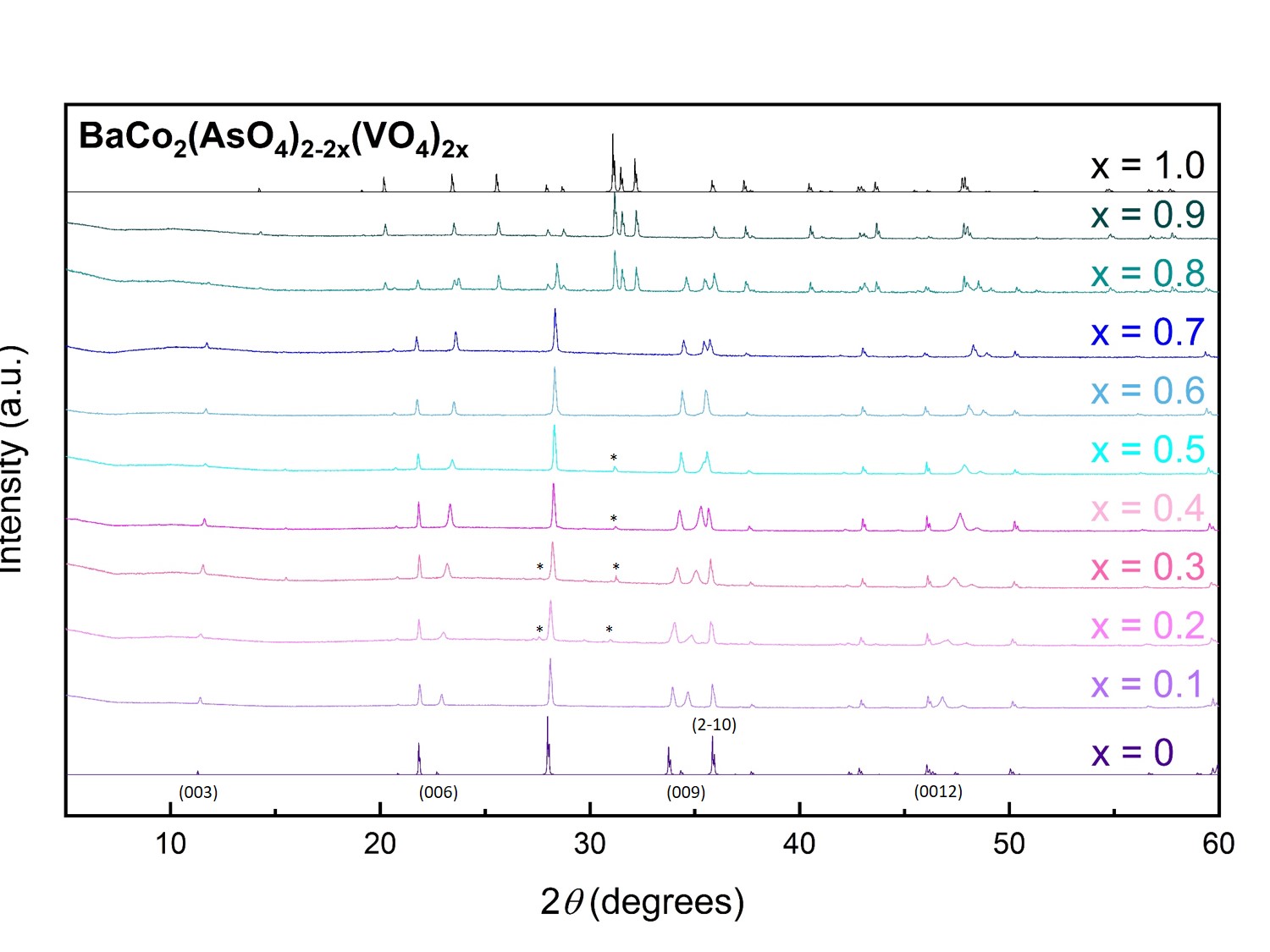}
  \caption{\textbf{pXRD patterns of representative \ch{BaCo2(AsO4)_{2-2x}(VO4)_{2x}} powder samples with 0~$\leq~x\leq$~1.0}. The simulated x~=~0 and x~=~1.0 patterns correspond to the BCAO and BCVO structures, respectively. Up to x~=~0.70, all reflections with an (00l) component shift rightward, indicative of a contraction of the unit cell along the c-axis. The sole (hk0) reflection shifts slightly leftward, indicative of a slight expansion of the unit cell in the \emph{ab} plane. Beyond x~=~0.70, the trigonal \emph{R$\overline{3}$H} (148) structure becomes less prevalent, leaving primarily the tetragonal structure of the parent \ch{BaCo2(VO4)2} phase (space group \emph{I41/acd} (142)). For both the x~=~0.80 and x~=~0.90 patterns, slight shifts are also observed in the reflections, likely due to partial As-substitution for V in the tetragonal structure. Reflections marked with an * correspond to the minor ($\sim$5 wt\%) \ch{Ba3(AsO4)2} impurity phase present in several compositions prior to secondary heating.}
  \label{pXRD}
\end{figure}
\newpage

\begin{landscape}

    \begin{table}[]
    \centering
    \begin{tabular}{|c|c|c|c|c|c|c|c|c|}
    \cline{3-8}
    \multicolumn{2}{c|}{\textbf{\ch{BaCo2(AsO4)_{2-2x}(VO4)_{2x}}}} & x = & 0 & 0.025 & 0.05 & 0.075 & 0.1\\ \hline
    & & a(\AA) & 5.0038(1) & 5.0069(1) & 5.0093(1) & 5.0089(1) & 5.0127(1)\\
    & & c(\AA) & 23.4908(7) & 23.4560(3) & 23.4225(5) & 23.4188(3) & 23.3453(6)\\
    Ba1 (\emph{3a}) & 0,0,0 & B$_{eq}$ & 0.2(2) & 0.10(5) & 0.63(8) & 0.27(7) & 0.95(12)\\
    Co1 (\emph{6c}) & 0,0,z & z & 0.1723(2) & 0.1708(1) & 0.1706(2) & 0.1704(2) & 0.1701(3)\\
    & & B$_{eq}$ & 0.5(2) & 0.21(7) & 0.49(11) & 0.69(8) & 0.46(14)\\
    As1/V1 (\emph{6c}) & $\frac{1}{3}$,$\frac{2}{3}$,z & z & 0.0910(2) & 0.0923(1) & 0.0919(1) & 0.0922(1) & 0.0921(1)\\
    & & As occ. & 1 & 0.980(10) & 0.956(19) & 0.889(12) & 0.900(20)\\
    & & V occ. & 0 & 0.020(10) & 0.044(19) & 0.111(12) & 0.100(20)\\
    & & B$_{eq}$ & 0.05(19) & 0.06(5) & 0.14(8) & 0.09(7) & 0.38(11)\\
    O1 (\emph{6c}) & $\frac{1}{3}$,$\frac{2}{3}$,z & z & 0.0203(10) & 0.0209(5) & 0.0218(5) & 0.0159(5) & 0.212(8)\\
    & & B$_{eq}$ & 2.2(4) & 1.22(15) & 0.9(2) & 0.16(16) & 1.4(3)\\
    O2 (\emph{18f}) & x,y,z & x & 0.0157(19) & 0.0096(14) & 0.0030(20) & 0.0010(30) & 0.0127(19)\\
    & & y & 0.3070(40) & 0.3230(20) & 0.3260(3) & 0.3180(3) & 0.3404(11)\\
    & & z & 0.1151(5) & 0.1162(3) & 0.1170(2) & 0.1150(1) & 0.1161(2)\\
    & & B$_{eq}$ & 2.2(4) & 1.22(15) & 0.9(2) & 0.16(16) & 1.4(3)\\
    & & R$_{wp}$ & 2.73 & 2.32 & 2.84 & 2.63 & 3.18\\
    & & R$_p$ & 2.02 & 1.80 & 2.19 & 2.06 & 2.18\\
    & & GOF & 1.99 & 1.28 & 1.47 & 1.25 & 1.69\\ \hline
    \end{tabular}
    \caption{\textbf{Results of Rietveld refinement to pXRD data collected on representative first-heat \ch{BaCo2(AsO4)_{2-2x}(VO4)_{2x}} powder samples (0 $\leq x\leq$ 0.1)}. All refinements were carried out in the \emph{R$\overline{3}$H} space group (148). All Ba1, Co1, O1/O2, and combined A1/V1 occupancies were fixed at 1, and the isotropic thermal parameters (B$_{eq}$) of As1/V1 and O1/O2 were set equal to one another by atom type.}
    \label{Rietveld_powders_H1_1}
\end{table}
\end{landscape}

\begin{landscape}
    \begin{table}[]
    \centering
    \begin{tabular}{|c|c|c|c|c|c|c|c|c|}
    \cline{3-8}
    \multicolumn{2}{c|}{\textbf{\ch{BaCo2(AsO4)_{2-2x}(VO4)_{2x}}}} & x = & 0.15 & 0.2 & 0.35 & 0.4 & 0.45\\ \hline
    & & a(\AA) & 5.0148(1) & 5.0203(2) & 5.0360(2) & 5.0407(2) & 5.0425(2)\\
    & & c(\AA) & 23.3223(5) & 23.2201(9) & 23.0039(7) & 22.9220(8) & 22.8685(7)\\
    Ba1 (\emph{3a}) & 0,0,0 & B$_{eq}$ & 0.20(11) & 0.92(14) & 1.81(11) & 0.9(6) & 0.93(11)\\
    Co1 (\emph{6c}) & 0,0,z & z & 0.1699(3) & 0.1706(3) & 0.1701(2) & 0.1700(3) & 0.1696(2)\\
    & & B$_{eq}$ & 0.06(12) & 0.48(15) & 0.47(12) & 0.4(5) & 0.49(14)\\
    As1/V1 (\emph{6c}) & $\frac{1}{3}$,$\frac{2}{3}$,z & z & 0.0923(1) & 0.0908(2) & 0.0910(1) & 0.0919(7) & 0.0905(2)\\
    & & As occ. & 0.816(17) & 0.83(2) & 0.650(15) & 0.550(110) & 0.550(20)\\
    & & V occ. & 0.184(17) & 0.17(2) & 0.350(15) & 0.450(110) & 0.450(20)\\
    & & B$_{eq}$ & 0.05(11) & 0.37(14) & 0.47(11) & 0.4(4) & 0.370(120)\\
    O1 (\emph{6c}) & $\frac{1}{3}$,$\frac{2}{3}$,z & z & 0.0155(8) & 0.0118(11) & 0.0146(5) & 0.0210(40) & 0.0149(7)\\
    & & B$_{eq}$ & 1.3(3) & 0.9(3) & 0.9(2) & 1.07(16) & 0.9(2)\\
    O2 (\emph{18f}) & x,y,z & x & 0.0100(40) & 0.0070(50) & 0.0160(30) & 0.0160(70) & 0.0190(30)\\
    & & y & 0.3343(10) & 0.3380(4) & 0.3380(20) & 0.3380(9) & 0.3440(40)\\
    & & z & 0.1135(2) & 0.1154(3) & 0.1148(2) & 0.1150(4) & 0.1145(2)\\
    & & B$_{eq}$ & 1.3(3) & 0.9(3) & 0.9(2) & 1.07(16) & 0.9(2)\\
    & & R$_{wp}$ & 3.27 & 3.05 & 3.02 & 2.55 & 3.06\\
    & & R$_p$ & 2.52 & 2.35 & 2.31 & 2.05 & 2.36\\
    & & GOF & 1.53 & 1.41 & 1.44 & 1.27 & 1.38\\ \hline
    \end{tabular}
    \caption{\textbf{Results of Rietveld refinement to pXRD data collected on representative first-heat \ch{BaCo2(AsO4)_{2-2x}(VO4)_{2x}} powder samples (0.15 $\leq x\leq$ 0.45)}. All refinements were carried out in the \emph{R$\overline{3}$H} space group (148). All Ba1, Co1, O1/O2, and combined A1/V1 occupancies were fixed at 1, and the isotropic thermal parameters (B$_{eq}$) of As1/V1 and O1/O2 were set equal to one another by atom type.}
    \label{Rietveld_powders_H1_2}
\end{table}
\end{landscape}

\begin{landscape}
    \begin{table}[]
    \centering
    \begin{tabular}{|c|c|c|c|c|c|c|c|c|}
    \cline{3-8}
    \multicolumn{2}{c|}{\textbf{\ch{BaCo2(AsO4)_{2-2x}(VO4)_{2x}}}} & x = & 0.025 & 0.05 & 0.075 & 0.1 & 0.15\\ \hline
    & & a(\AA) & 5.0066(1) & 5.0082(1) & 5.0090(1) & 5.0122(1) & 5.0148(1)\\
    & & c(\AA) & 23.4476(1) & 23.4184(4) & 23.4105(4) & 23.3563(6) & 23.3328(6)\\
    Ba1 (\emph{3a}) & 0,0,0 & B$_{eq}$ & 0.64(10) & 0.92(9) & 1.03(8) & 0.92(10) & 0.95(9)\\
    Co1 (\emph{6c}) & 0,0,z & z & 0.1688(2) & 0.1702(2) & 0.1695(2) & 0.1705(2) & 0.1702(2)\\
    & & B$_{eq}$ & 0.40(10) & 0.49(11) & 0.59(10) & 0.48(11) & 0.49(11)\\
    As1/V1 (\emph{6c}) & $\frac{1}{3}$,$\frac{2}{3}$,z & z & 0.0914(1) & 0.0920(1) & 0.0925 (1) & 0.0920(1) & 0.0919(1)\\
    & & As occ. & 0.976(15) & 0.934(17) & 0.960(13) & 0.900(15) & 0.850(15)\\
    & & V occ. & 0.024(15) & 0.066(17) & 0.040(13) & 0.100(15) & 0.150(15)\\
    & & B$_{eq}$ & 0.38(9) & 0.40(9) & 0.42(8) & 0.37(9) & 0.37(9)\\
    O1 (\emph{6c}) & $\frac{1}{3}$,$\frac{2}{3}$,z & z & 0.0174(6) & 0.3138(5) & 0.0200(4) & 0.0135(5) & 0.0132(5)\\
    & & B$_{eq}$ & 0.1(2) & 0.97(18) & 0.93(17) & 0.9(2) & 0.9(2)\\
    O2 (\emph{18f}) & x,y,z & x & 0.0170(30) & 0.0160(20) & 0.0170(2) & 0.0100(2) & 0.0120(2)\\
    & & y & 0.3376(9) & 0.3378(15) & 0.3346(19) & 0.3380(4) & 0.3380(2)\\
    & & z & 0.1159(2) & 0.1170(2) & 0.1161(1) & 0.1159(2) & 0.1148(2)\\
    & & B$_{eq}$ & 0.1(2) & 0.97(18) & 0.93(17) & 0.9(2) & 0.9(2)\\
    & & R$_{wp}$ & 3.19 & 2.96 & 2.94 & 2.86 & 2.79\\
    & & R$_p$ & 2.55 & 2.32 & 2.28 & 2.27 & 2.17\\
    & & GOF & 1.49 & 1.37 & 1.40 & 1.28 & 1.31\\ \hline
    \end{tabular}
    \caption{\textbf{Results of Rietveld refinement to pXRD data collected on representative second-heat \ch{BaCo2(AsO4)_{2-2x}(VO4)_{2x}} crystal samples (0.025 $\leq x\leq$ 0.15)}. All refinements were carried out in the \emph{R$\overline{3}$H} space group (148). All Ba1, Co1, O1/O2, and combined A1/V1 occupancies were fixed at 1, and the isotropic thermal parameters (B$_{eq}$) of As1/V1 and O1/O2 were set equal to one another by atom type.}
    \label{Rietveld_powders_H2_1}
\end{table}
\end{landscape}

\begin{landscape}
    \begin{table}[]
    \centering
    \begin{tabular}{|c|c|c|c|c|c|c|c|c|}
    \cline{3-8}
    \multicolumn{2}{c|}{\textbf{\ch{BaCo2(AsO4)_{2-2x}(VO4)_{2x}}}} & x = & 0.2 & 0.25 & 0.3 & 0.5 & 0.7\\ \hline
    & & a(\AA) & 5.0210(2) & 5.0322(2) & 5.0328(2) & 5.0373(1) & 5.0577(2)\\
    & & c(\AA) & 23.2152(9) & 23.1546(5) & 23.0535(11) & 22.9671(5) & 22.6559(7)\\
    Ba1 (\emph{3a}) & 0,0,0 & B$_{eq}$ & 0.93(12) & 2.00(13) & 0.93(13) & 0.91(10) & 0.92(14)\\
    Co1 (\emph{6c}) & 0,0,z & z & 0.1698(3) & 0.1707(2) & 0.1713(3) & 0.1707(2) & 0.1702(3)\\
    & & B$_{eq}$ & 0.78(14) & 1.91(11) & 0.48(14) & 0.63(11) & 0.48(16)\\
    As1/V1 (\emph{6c}) & $\frac{1}{3}$,$\frac{2}{3}$,z & z & 0.0918(1) & 0.0912(1) & 0.0919(1) & 0.0904(1) & 0.0909(2)\\
    & & As occ. & 0.780(20) & 0.719(12) & 0.696(16) & 0.581(15) & 0.401(18)\\
    & & V occ. & 0.220(20) & 0.281(12) & 0.304(16) & 0.419(15) & 0.599(18)\\
    & & B$_{eq}$ & 0.43(13) & 1.08(11) & 0.37(14) & 0.38(11) & 0.47(14)\\
    O1 (\emph{6c}) & $\frac{1}{3}$,$\frac{2}{3}$,z & z & 0.0188(6) & 0.0033(9) & 0.0160(8) & 0.0100(9) & 0.0207(11)\\
    & & B$_{eq}$ & 0.9(2) & 0.77(17) & 0.9(3) & 0.9(2) & 0.8(3)\\
    O2 (\emph{18f}) & x,y,z & x & 0.0160(3) & 0.0540(50) & 0.0250(40) & 0.0160(30) & 0.0160(40)\\
    & & y & 0.3370(2) & 0.3600(40) & 0.3404(11) & 0.3404(15) & 0.3375(14)\\
    & & z & 0.1158(2) & 0.1114(1) & 0.11520(2) & 0.1131(2) & 0.1096(2)\\
    & & B$_{eq}$ & 0.9(2) & 0.77(17) & 0.9(3) & 0.9(2) & 0.8(3)\\
    & & R$_{wp}$ & 3.01 & 3.49 & 3.75 & 3.01 & 2.99\\
    & & R$_p$ & 2.37 & 2.51 & 2.74 & 2.33 & 2.34\\
    & & GOF & 1.30 & 1.51 & 1.63 & 1.35 & 1.30\\ \hline
    \end{tabular}
    \caption{\textbf{Results of Rietveld refinement to pXRD data collected on representative second-heat \ch{BaCo2(AsO4)_{2-2x}(VO4)_{2x}} crystal samples (0.2 $\leq x\leq$ 0.7)}. All refinements were carried out in the \emph{R$\overline{3}$H} space group (148). All Ba1, Co1, O1/O2, and combined A1/V1 occupancies were fixed at 1, and the isotropic thermal parameters (B$_{eq}$) of As1/V1 and O1/O2 were set equal to one another by atom type.}
    \label{Rietveld_powders_H2_2}
\end{table}
\end{landscape}

\begin{landscape}
\subsection{SCXRD refinement data}
\begin{table}
    \caption{\textbf{BCAO-V Single Crystal and Structural Refinement Data}}
    \label{SC Data 1a}
   \begin{tabular}{|l|l|l|l|l|l|}
   \cline{2-6}
   \multicolumn{1}{c|}{\textbf{\ch{BaCo2(AsO4)_{2-2x}(VO4)_{2x}}}} 
    & x=0.025 & 0.05 & 0.075 & 0.1 & 0.5 \\ \hline
    Formula weight (g $mol^{-1}$) & 531.83 & 530.64 & 529.44 & 528.23 & 509.05\\
    Crystal system & trigonal & trigonal & trigonal & trigonal & trigonal\\
    Space group & \emph{R$\overline{3}$H} (148) & \emph{R$\overline{3}$H} (148) & \emph{R$\overline{3}$H} (148) & \emph{R$\overline{3}$H} (148) & \emph{R$\overline{3}$H} (148)\\
    \emph{a}(\AA{}) &  5.0008(2) & 5.0052(2) & 5.0089(2) & 5.0180(2) & 5.0353(2)\\
    \emph{b}(\AA{}) &  5.0008(2) & 5.0052(2) & 5.0089(2) & 5.0180(2) & 5.0353(2)\\
    \emph{c}(\AA{}) &  23.3450(10) & 23.3259(9) & 23.2848(9) & 23.2660(10) & 22.9252(9)\\ 
    Volume (\AA{}$^3$) & 505.60(5) & 506.07(4) & 505.93(4) & 507.36(5) & 503.38(4)\\
    \emph{Z} & 1 & 1 & 1 & 1 & 1\\
    Radiation & Mo \emph{K}$\alpha$ & Mo \emph{K}$\alpha$ & Mo \emph{K}$\alpha$ & Mo \emph{K}$\alpha$ & Mo \emph{K}$\alpha$\\
    & ($\lambda$ = 0.71073 \AA{}) & ($\lambda$ = 0.71073 \AA{}) & ($\lambda$ = 0.71073 \AA{}) & ($\lambda$ = 0.71073 \AA{}) & ($\lambda$ = 0.71073 \AA{}) \\
    Temperature (K) & 213(2) & 213(2) & 213(2) & 213(2) & 213(2)\\
    Reflections collected/unique & 4505/438 & 4570/438 & 4546/438 & 4576/438 & 5184/436\\
    $R_{int}$ & 0.0450 & 0.0358 & 0.0369 & 0.0384 & 0.0350\\
    Data/parameters & 438/23 & 438/23 & 438/23 & 438/23 & 436/23\\
    Goodness-of-fit &  1.149 & 1.196 & 1.166 & 1.136 & 1.179\\
    $R_1$ [$F^2$ $>$ $2\sigma(F^2)$];$^a$ $R_1$ [all data] & 0.0165; 0.0169 & 0.0147; 0.0156 & 0.0144; 0.0160 & 0.0128; 0.0150 & 0.0135; 0.0146\\
    $wR(F^2)$$^b$ & 0.0417 & 0.0384 & 0.0366 & 0.0310 & 0.0347\\
    Largest diff. peak and hole (e \AA{}$^{-3}$) & 0.830 and -2.055 & 0.700 and -2.189 & 0.988 and -1.320 & 0.669 and -0.692 & 0.752 and -0.708\\ \hline
    \multicolumn{1}{l}{$^a$R(F)=$\Sigma$$|$$|$F$_o$$|$-$|$F$_c$$|$$|$/$\Sigma$$|$$|$F$_o$$|$}\\
    \multicolumn{1}{l}{$^b$R$_\omega$(F$^2_o$)= [$\Sigma\omega$(F$^2_o$-F$^2_c$)$^2$/$\Sigma\omega$(F$^2_o$)$^2$]$^{1/2}$}
    \end{tabular} 
\end{table}
\end{landscape}

\begin{landscape}
    \begin{table}
    \caption{\textbf{BCAO-V (x~=~0.025) Atomic Positions and Anisotropic Displacement Parameters}. All $U_{ij}$ values are given in units of $\textrm\AA{^2}$.}
    \label{SC Data 2.5}
    \centering
    \begin{tabular}{|c|c|c|c|c|c|c|c|c|c|c|}
    \hline
       Atom & Wyckoff Site & x & y & z & $U_{11}$ & $U_{22}$ & $U_{33}$ & $U_{23}$ & $U_{13}$ & $U_{12}$\\ \hline
       Ba1 & 3a & 0 & 0 & 0 & 0.00893(11) & 0.00893(11) & 0.00860(16) & 0 & 0 & 0.00446(6)\\
       Co1 & 6c & 0 & 0 & 0.17023(2) & 0.00495(13) & 0.00495(13) & 0.0068(2) & 0 & 0 & 0.00248(7)\\
       As1 & 6c & 1/3 & 2/3 & 0.09162(2) & 0.00432(13) & 0.00432(13) & 0.00490(18) & 0 & 0 & 0.00216(7)\\
       O1 & 6c & 1/3 & 2/3 & 0.02082(11) & 0.0120(6) & 0.0120(6) & 0.0070(11) & 0 & 0 & 0.0060(3)\\
       O2 & 18f & 0.0164(3) & 0.3370(3) & 0.11460(6) & 0.0057(5) & 0.0055(5) & 0.0090(6) & 0.0017(5) & 0.0013(5) & 0.0019(4)\\ 
       \hline
    \end{tabular}
\end{table} 

\begin{table}
    \caption{\textbf{BCAO-V (x~=~0.05) Atomic Positions and Anisotropic Displacement Parameters}. All $U_{ij}$ values are given in units of $\textrm\AA{^2}$.}
    \label{SC Data 5}
    \centering
    \begin{tabular}{|c|c|c|c|c|c|c|c|c|c|c|}
    \hline
       Atom & Wyckoff Site & x & y & z & $U_{11}$ & $U_{22}$ & $U_{33}$ & $U_{12}$ & $U_{13}$ & $U_{23}$\\ \hline
       Ba1 & 3a & 0 & 0 & 0 & 0.00871(11) & 0.00871(11) & 0.00913(14) & 0 & 0 & 0.00435(5)\\
       Co1 & 6c & 0 & 0 & 0.17026(2) & 0.00479(13) & 0.00479(13) & 0.00745(19) & 0 & 0 & 0.00240(6)\\
       As1 & 6c & 1/3 & 2/3 & 0.09163(2) & 0.00429(13) & 0.00429(13) & 0.00548(17) & 0 & 0 & 0.00215(6)\\
       O1 & 6c & 1/3 & 2/3 & 0.02073(10) & 0.0106(6) & 0.0106(6) & 0.0092(10) & 0 & 0 & 0.0053(3)\\
       O2 & 18f & 0.0160(3) & 0.3366(3) & 0.11456(5) & 0.0054(5) & 0.0061(5) & 0.0097(6) & 0.0015(4) & 0.0011(4) & 0.0023(4)\\ 
       \hline
    \end{tabular}
\end{table}
\end{landscape}

\begin{landscape}
\begin{table}
    \caption{\textbf{BCAO-V (x~=~0.075) Atomic Positions and Anisotropic Displacement Parameters}. All $U_{ij}$ values are given in units of $\textrm\AA{^2}$.}
    \label{SC Data 7.5}
    \centering
    \begin{tabular}{|c|c|c|c|c|c|c|c|c|c|c|}
    \hline
       Atom & Wyckoff Site & x & y & z & $U_{11}$ & $U_{22}$ & $U_{33}$ & $U_{12}$ & $U_{13}$ & $U_{23}$\\ \hline
       Ba1 & 3a & 0 & 0 & 0 & 0.00891(11) & 0.00891(11) & 0.01042(15) & 0 & 0 & 0.00445(5)\\
       Co1 & 6c & 0 & 0 & 0.17026(2) & 0.00516(13) & 0.00516(13) & 0.0089(2) & 0 & 0 & 0.00258(7)\\
       As1 & 6c & 1/3 & 2/3 & 0.09156(2) & 0.00446(13) & 0.00446(13) & 0.00667(18) & 0 & 0 & 0.00223(7)\\
       O1 & 6c & 1/3 & 2/3 & 0.02027(11) & 0.0098(7) & 0.0098(7) & 0.0093(11) & 0 & 0 & 0.0049(3)\\
       O2 & 18f & 0.0162(3) & 0.3368(3) & 0.11452(6) & 0.0054(6) & 0.0052(6) & 0.0108(6) & 0.0022(5) & 0.0013(5) & 0.0024(5)\\ 
       \hline
    \end{tabular}
\end{table}

\begin{table}
    \caption{\textbf{BCAO-V (x~=~0.1) Atomic Positions and Anisotropic Displacement Parameters}. All $U_{ij}$ values are given in units of $\textrm\AA{^2}$.}
    \label{SC Data 10}
    \centering
    \begin{tabular}{|c|c|c|c|c|c|c|c|c|c|c|}
    \hline
       Atom & Wyckoff Site & x & y & z & $U_{11}$ & $U_{22}$ & $U_{33}$ & $U_{12}$ & $U_{13}$ & $U_{23}$\\ \hline
       Ba1 & 3a & 0 & 0 & 0 & 0.00883(10) & 0.00883(10) & 0.00997(13) & 0 & 0 & 0.00441(5)\\
       Co1 & 6c & 0 & 0 & 0.17035(2) & 0.00509(12) & 0.00509(12) & 0.00818(19) & 0 & 0 & 0.00254(6)\\
       As1 & 6c & 1/3 & 2/3 & 0.09147(2) & 0.00421(13) & 0.00421(13) & 0.00636(16) & 0 & 0 & 0.00210(6)\\
       O1 & 6c & 1/3 & 2/3 & 0.01997(9) & 0.0098(7) & 0.0098(7) & 0.0075(10) & 0 & 0 & 0.0049(3)\\
       O2 & 18f & 0.0161(3) & 0.3367(3) & 0.11442(5) & 0.0056(6) & 0.0049(5) & 0.0112(6) & 0.0016(5) & 0.0013(4) & 0.0020(5)\\ 
       \hline
    \end{tabular}
\end{table}
\end{landscape}

\begin{landscape}
\begin{table}
    \caption{\textbf{BCAO-V (x~=~0.5) Atomic Positions and Anisotropic Displacement Parameters}. All $U_{ij}$ values are given in units of $\textrm\AA{^2}$.}
    \label{SC Data 50}
    \centering
    \begin{tabular}{|c|c|c|c|c|c|c|c|c|c|c|}
    \hline
       Atom & Wyckoff Site & x & y & z & $U_{11}$ & $U_{22}$ & $U_{33}$ & $U_{12}$ & $U_{13}$ & $U_{23}$\\ \hline
       Ba1 & 3a & 0 & 0 & 0 & 0.00958(10) & 0.00958(10) & 0.00927(13) & 0 & 0 & 0.00479(5)\\
       Co1 & 6c & 0 & 0 & 0.17065(2) & 0.00537(12) & 0.00537(12) & 0.00824(18) & 0 & 0 & 0.00268(6)\\
       As1 & 6c & 1/3 & 2/3 & 0.09099(2) & 0.00507(14) & 0.00507(14) & 0.00630(17) & 0 & 0 & 0.00253(7)\\
       O1 & 6c & 1/3 & 2/3 & 0.01845(9) & 0.0124(6) & 0.0124(6) & 0.0069(9) & 0 & 0 & 0.0062(3)\\
       O2 & 18f & 0.0156(3) & 0.3353(3) & 0.11385(5) & 0.0066(5) & 0.0071(5) & 0.0111(5) & 0.0022(4) & 0.0014(4) & 0.0029(4)\\ 
       \hline
    \end{tabular}
\end{table}
\end{landscape} 

\begin{figure}
  \includegraphics[width=1.0\textwidth]
  {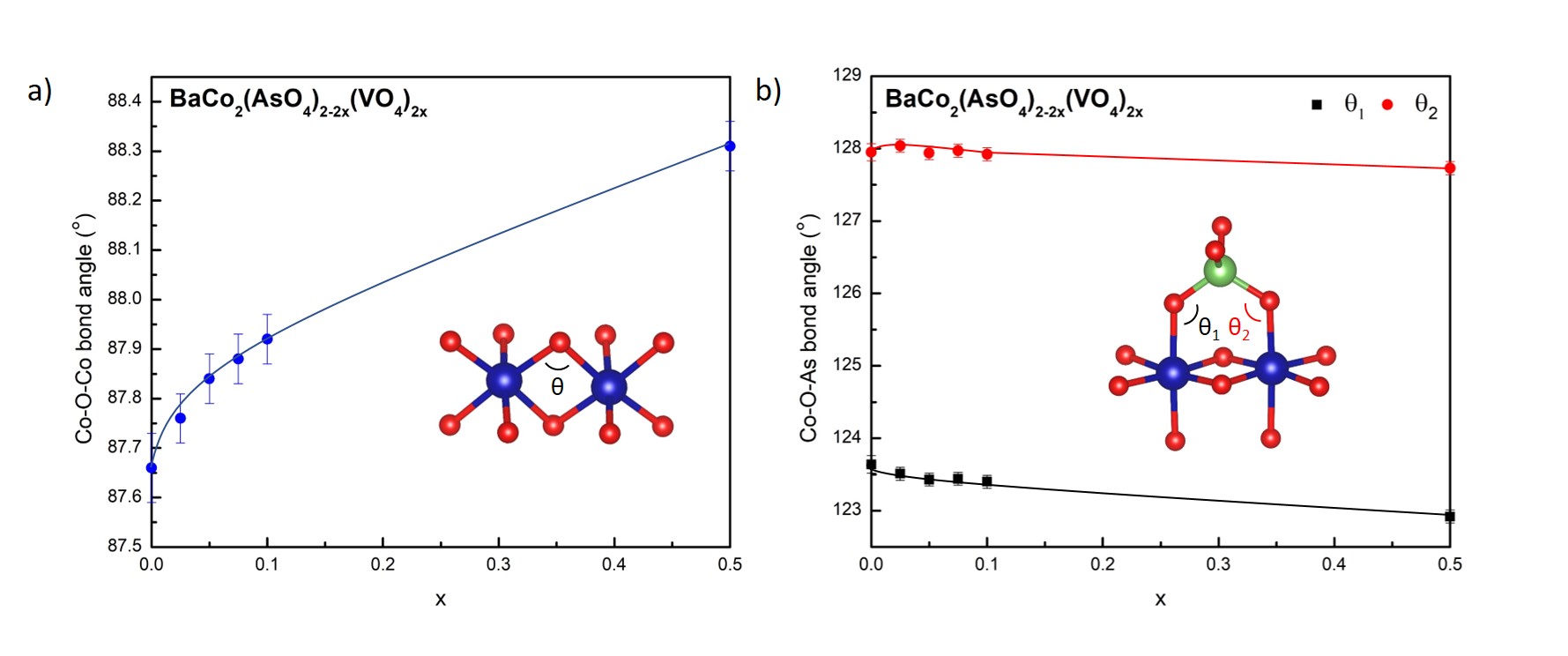}
  \caption{\textbf{Changes in a) Co-O-Co and b) Co-O-As bond angles in BCAO as a function of vanadium substitution.} As more vanadium enters the structure, the Co-O-Co angle is observed to increase slightly, indicating a weakening of the FM $J_1$ superexchange along this pathway. Increasing substitution decreases the Co-O-As bond angles in the structure, which reduces orbital overlap and is expected to decrease the magnitude of the AFM $J_3$ exchange. The partial replacement of the strongly covalently-bonded arsenate group with more ionically-bonded vanadate also decreases orbital overlap along this exchange pathway, further reducing the magnitude of $J_3$. The changes in both $J_1$ and $J_3$ as a function of vanadium substitution indicate a gradual tuning of the magnetic ground state, through to the critical point observed at around x~=~0.10. All values were derived from single crystal diffraction data of the compositions shown.}
  \label{J1_J3_SC}
\end{figure} \newpage

\begin{table}
    \caption{{\textbf{Selected bond angles of nominally-substituted BCAO-V compositions.} Upon increasing x, the Co-O-Co bridging angle between neighboring Co atoms is observed to gradually increase. The strength of FM superexchange along this pathway and thus the magnitude of $J_1$ is expected to decrease. With increasing x, the Co-O-As bridging angle is observed to decrease, also reducing the magnitude of the AFM $J_3$ exchange.}}
    \label{SC_derived_params}
    \centering
    \begin{tabular}{|c|c|c|}
    \hline
     Nominal V-content & Co-O-Co bond angle & Co-O-As bond angles \\ 
     (x~=~) & (degrees) & (degrees)\\ \hline
     0 & 87.66(7) & 123.64(12), 127.95(12)\\
     0.025 & 87.76(5) & 123.51(9), 128.04(9)\\
     0.05 & 87.84(5) & 123.43(9), 127.94(9)\\
     0.075 & 87.88(5) & 123.44(9), 127.97(9)\\
     0.10 & 87.92(5) & 123.40(9), 127.92(9)\\
     0.50 & 88.31(5) & 122.92(9), 127.73(9)\\ \hline
    \end{tabular}
\end{table} \newpage

\subsection{Laue diffraction data}
\begin{figure}
  \includegraphics[width=1.0\textwidth]
  {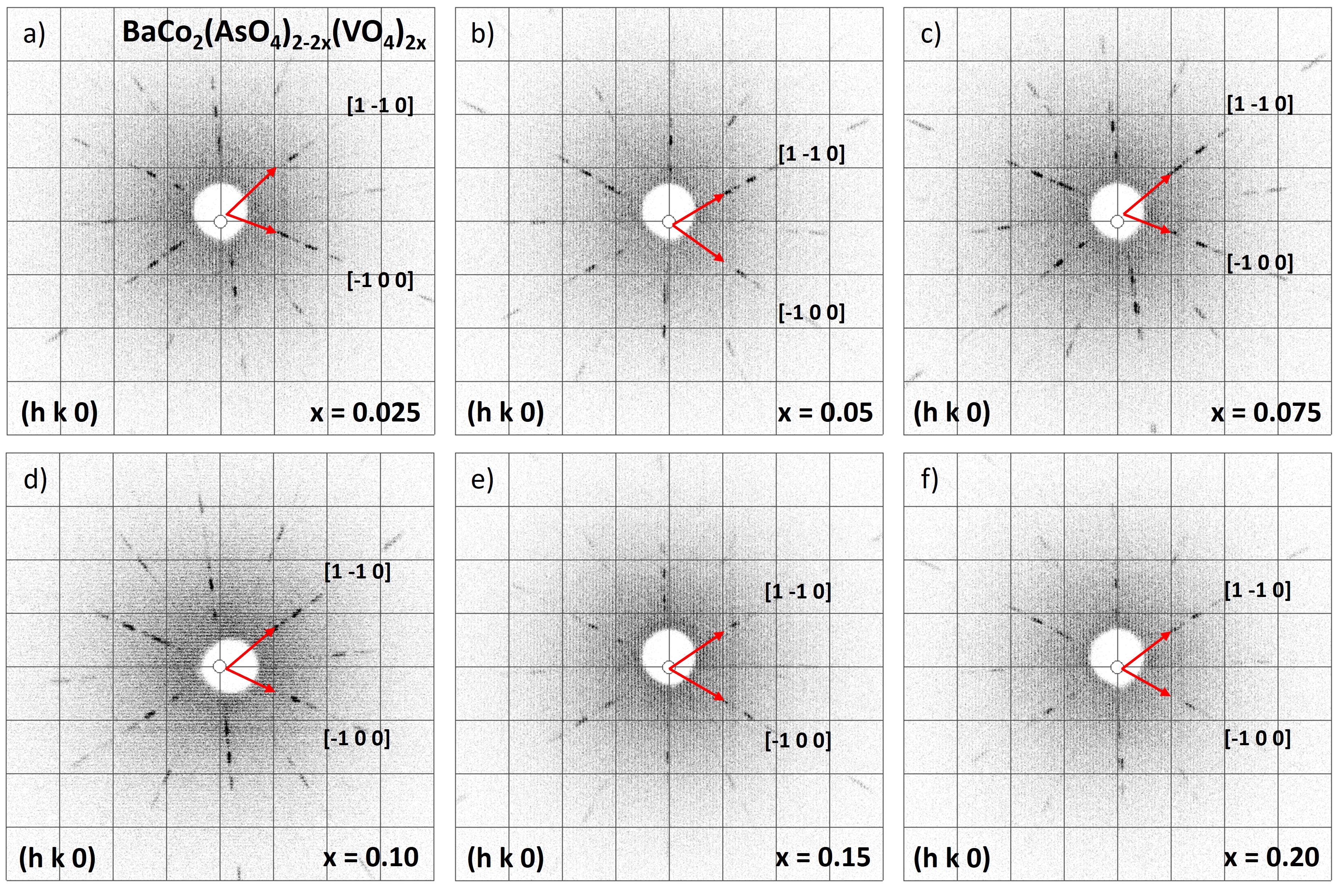}
  \caption{\textbf{Laue diffraction patterns of representative a) x~=~0.025, b) x~=~0.05, c) x~=~0.075, d) x~=~0.10, e) x~=~0.15, and f) x~=~0.20 BCAO-V single crystals in the (h~k~0) plane.} The [1 -1 0] and [-1 0 0] directions are denoted by the red arrows.}
  \label{Laue}
\end{figure} \newpage

\subsection{Raman scattering spectra}
\begin{figure}
  \includegraphics[width=0.65\textwidth]
  {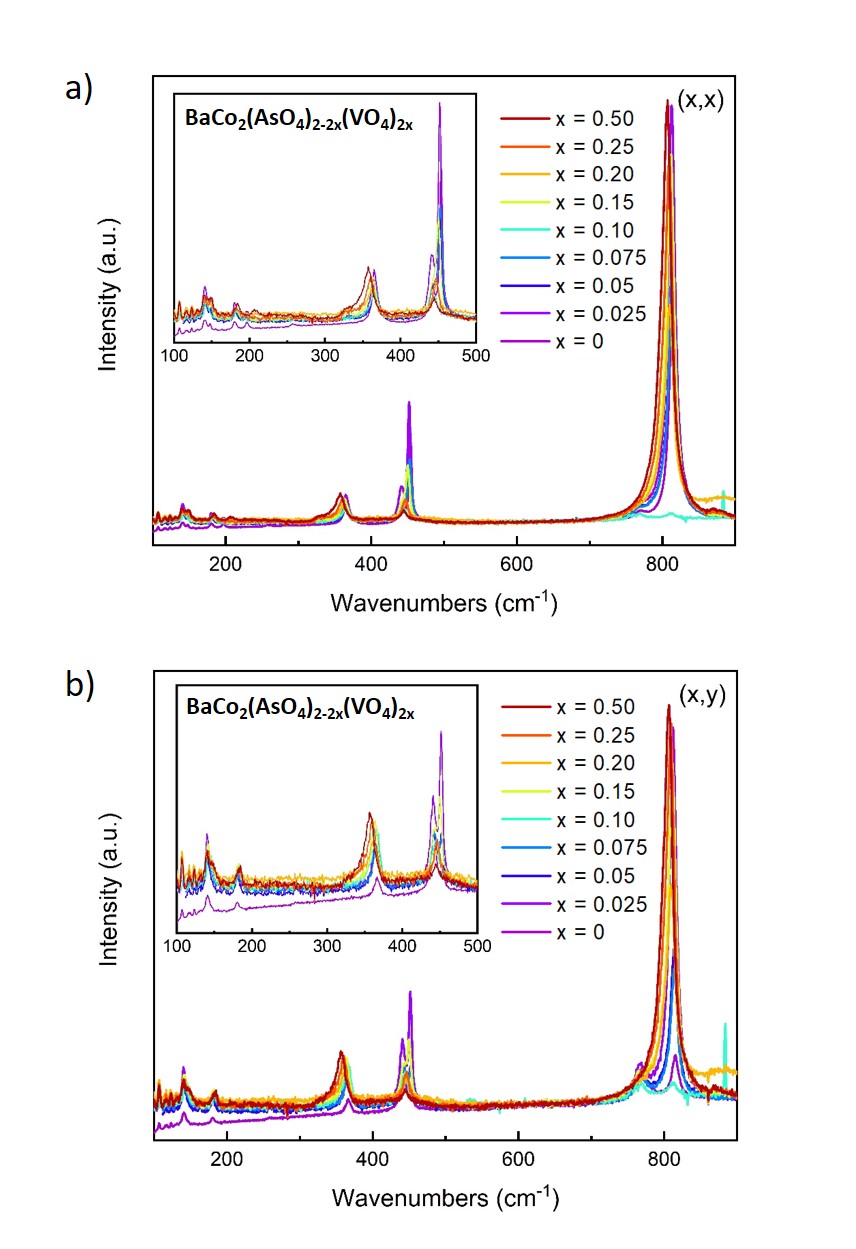}
  \caption{\textbf{Raman scattering spectra of BCAO-V compositions in the a) (x,x) and b) (x,y) polarization configurations as a function of V-substitution.} Enlarged view of the lower-frequency vibrational modes (insets).}
  \label{Raman_xx_xy_sep}
\end{figure} \newpage

\begin{figure}
  \includegraphics[width=1.00\textwidth]
  {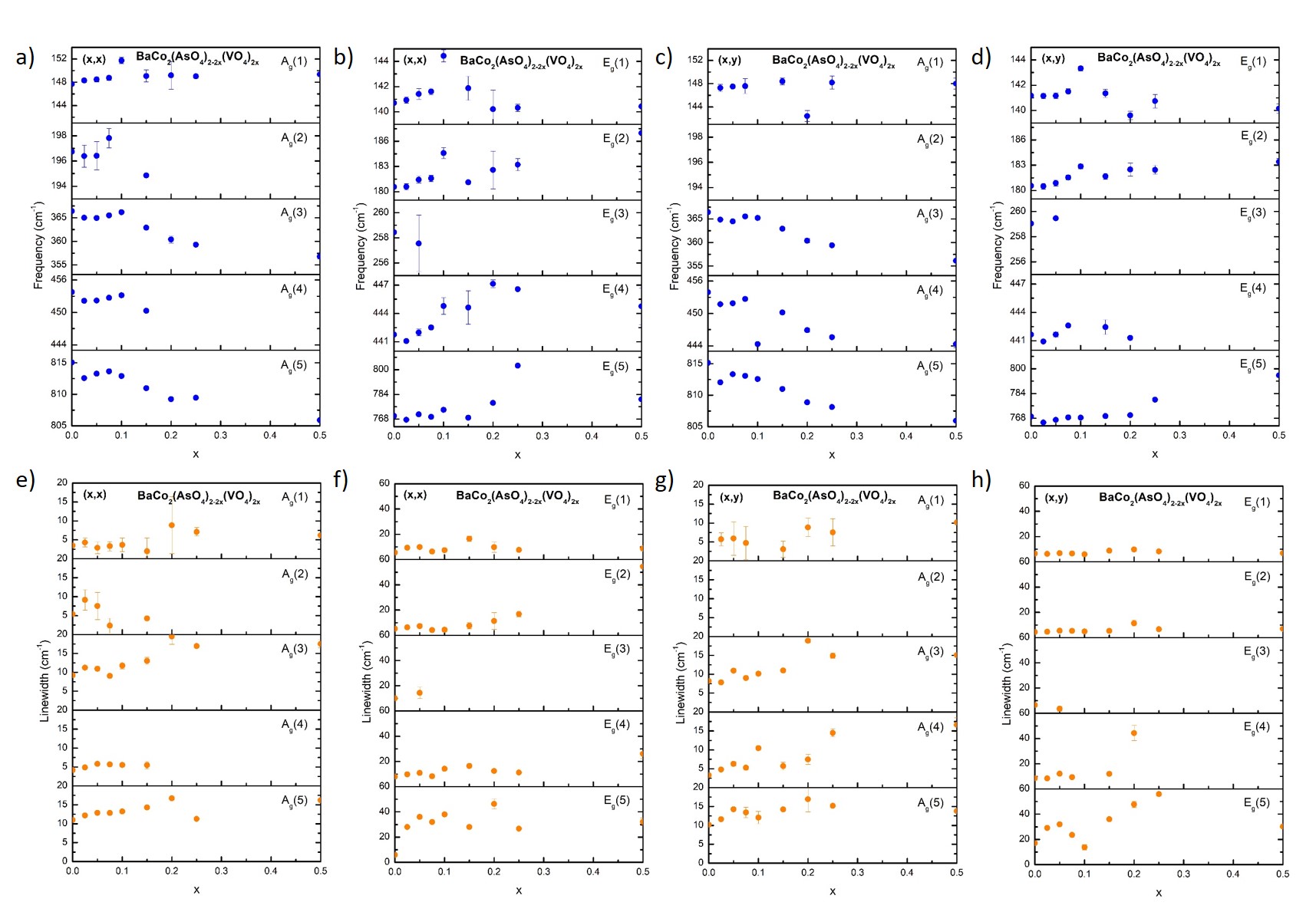}
  \caption{\textbf{Measured Raman frequency shifts of singlet A$_g$ (a, c) and doublet E$_g$ (b, d) modes as a function of temperature with unpolarized (x,x) and polarized (x,y) light, respectively (blue).} Measured linewidths of singlet A$_g$ (e, g) and doublet E$_g$ (f, h) modes as a function of temperature in the (x,x) and (x,y) polarization configurations, respectively (orange).}
  \label{Raman2}
\end{figure} \newpage

\begin{figure}
  \includegraphics[width=1.0\textwidth]
  {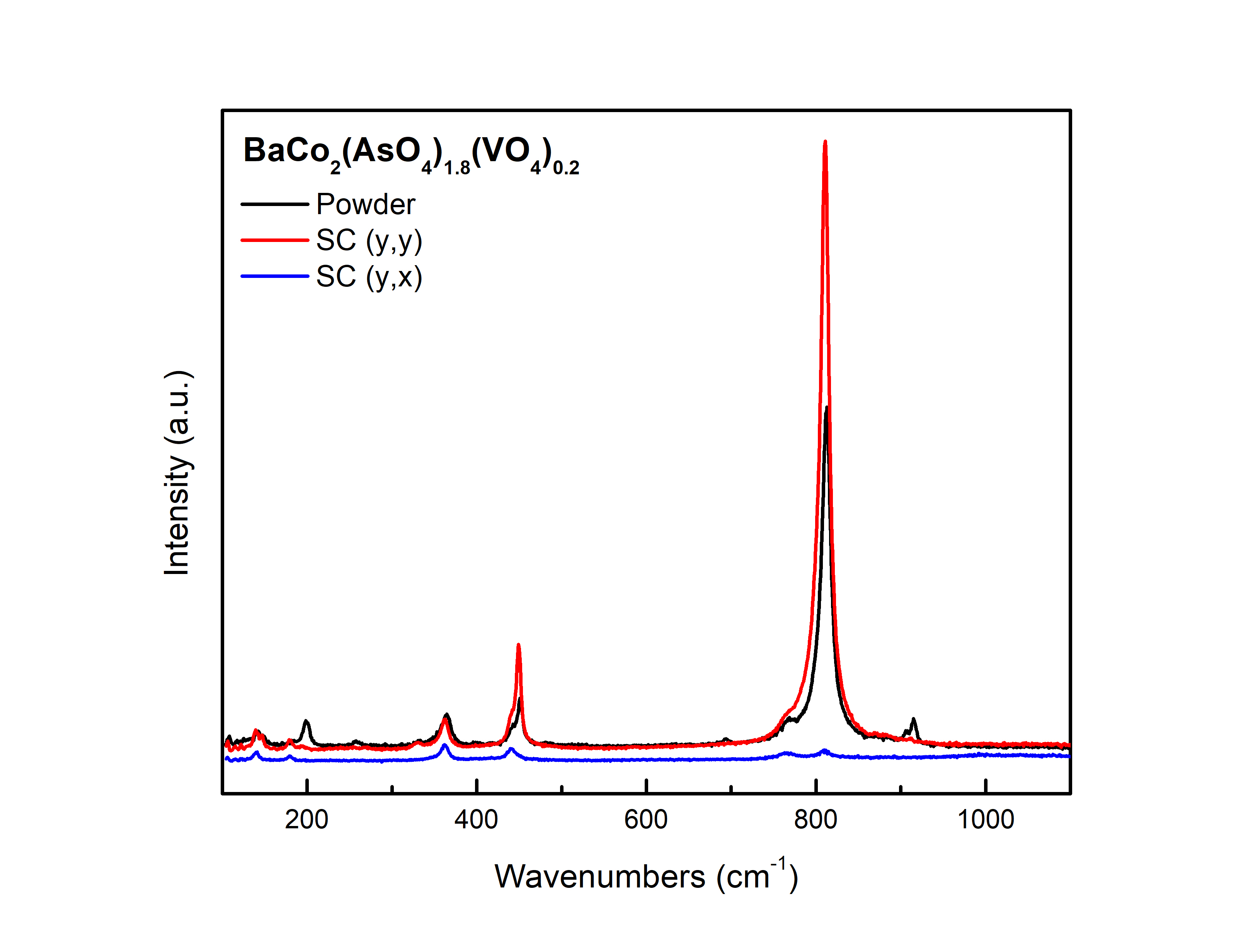}
  \caption{\textbf{Raman scattering spectra of \ch{BaCo2(AsO4)_{1.8}(VO4)_{0.2}} powder (black) and single crystal compositions in the (red) (x,x) and (blue) (x,y) polarization configurations.}}
  \label{Raman_10p0_powder_SC}
\end{figure} \newpage

\begin{figure}
  \includegraphics[width=1.00\textwidth]
  {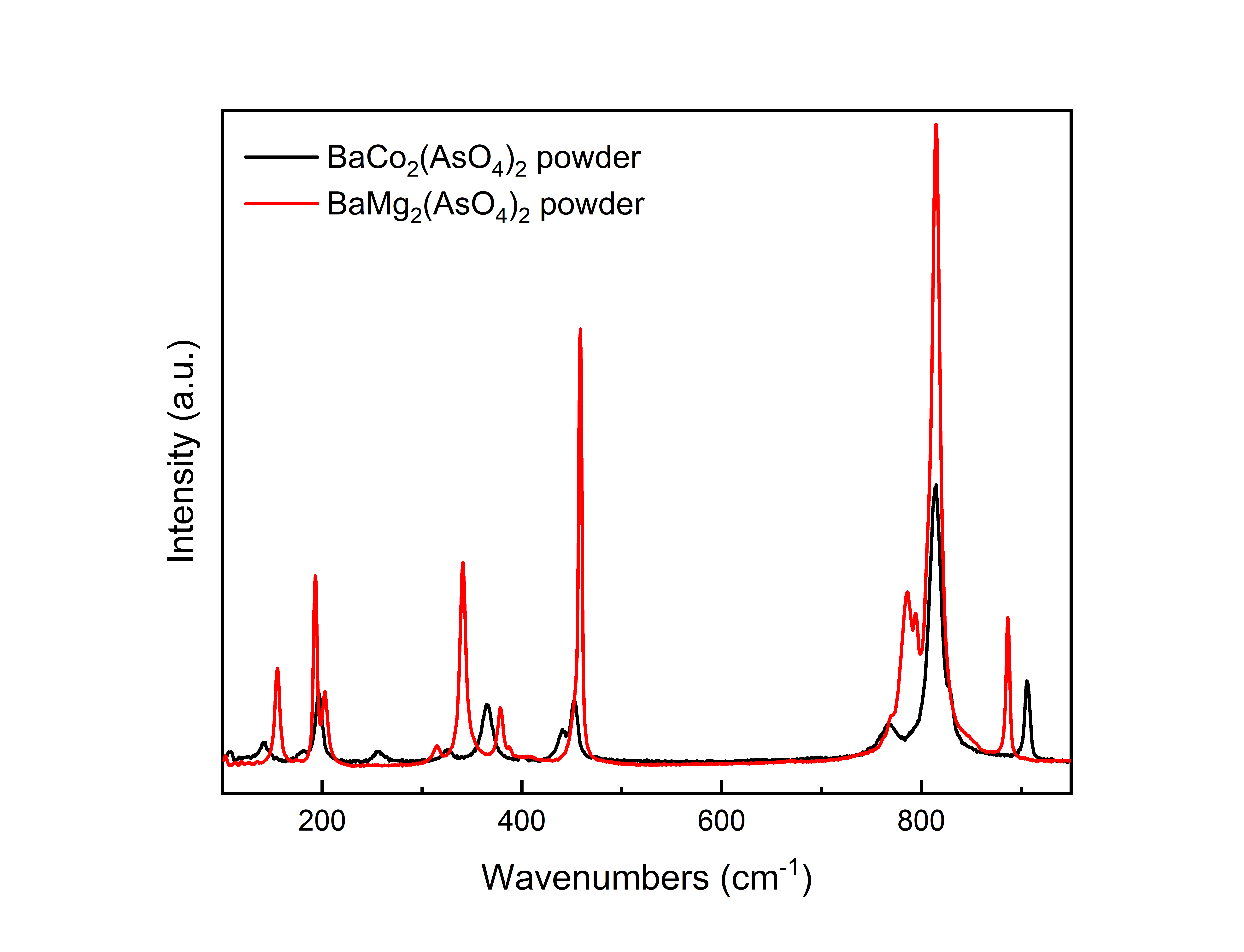}
  \caption{\textbf{Raman scattering spectra collected from BCAO (black) and BMAO powder samples (red) in the (x,x) and (y,y) polarization configurations, respectively.} While the two phases share the same structure, the two spectra differ by more than a uniform frequency shift, supporting that the vibrational modes in each differ considerably.}
  \label{Raman_mag_non-mag}
\end{figure}
\newpage

\section{Compositional analysis}
\begin{table}
    \caption{\textbf{ICP-OES-derived vanadium content of nominally-substituted BCAO-V compositions.} All measurements were performed in triplicate. * denotes a second-heat x~=~0.10 crystal sample grown via the Bridgman method, whereas the other samples were grown from the melt with no particular directional temperature gradient.}
    \label{ICP table}
    \centering
    \begin{tabular}{|c|c|c|c|c|c|c|c|c|c|c|}
    \hline
     Nominal V-content & Sample State & As & Ba & Co & V & V-content\\ 
     (x~=~) & & (wt\%) & (wt\%) & (wt\%) & (wt\%) & (x~=~) \\ \hline
     0.025 & crystal & 25.6(7) & 23.1(9) & 19.8(4) & 0.50(1) & 0.029(1) \\
     0.05 & crystal & 25.8(2) & 23.3(7) & 20.1(6) & 1.03(1) & 0.059(1) \\
     0.075 & crystal & 25.8(7) & 24.2(3) & 20.6(1.0) & 1.43(3) & 0.082(1) \\
     0.10 & crystal & 25.3(8) & 22.3(1.4) & 21.1(1.3) & 1.81(1) & 0.105(1) \\
     0.10 & powder & 24.5(6) & 23.4(4) & 21.6(4) & 1.87(7) & 0.112(1) \\
     0.10* & crystal & 23.2(3) & 24.0(7) & 20.8(0.4) & 2.75(4) & 0.175(2) \\
     0.15 & crystal & 24.3(3) & 21.4(1.9) & 20.2(4) & 2.63(4) & 0.160(2) \\
     0.20 & crystal & 22.4(3) & 23.5(5) & 20.9(5) & 2.71(1) & 0.244(2) \\
     0.70 & crystal & 8.88(16) & 24.5(1) & 22.0(2.8) & 13.6(2) & 0.557(13) \\ \hline
    \end{tabular}
\end{table} \newpage

\begin{figure}
  \includegraphics[width=1.00\textwidth]
  {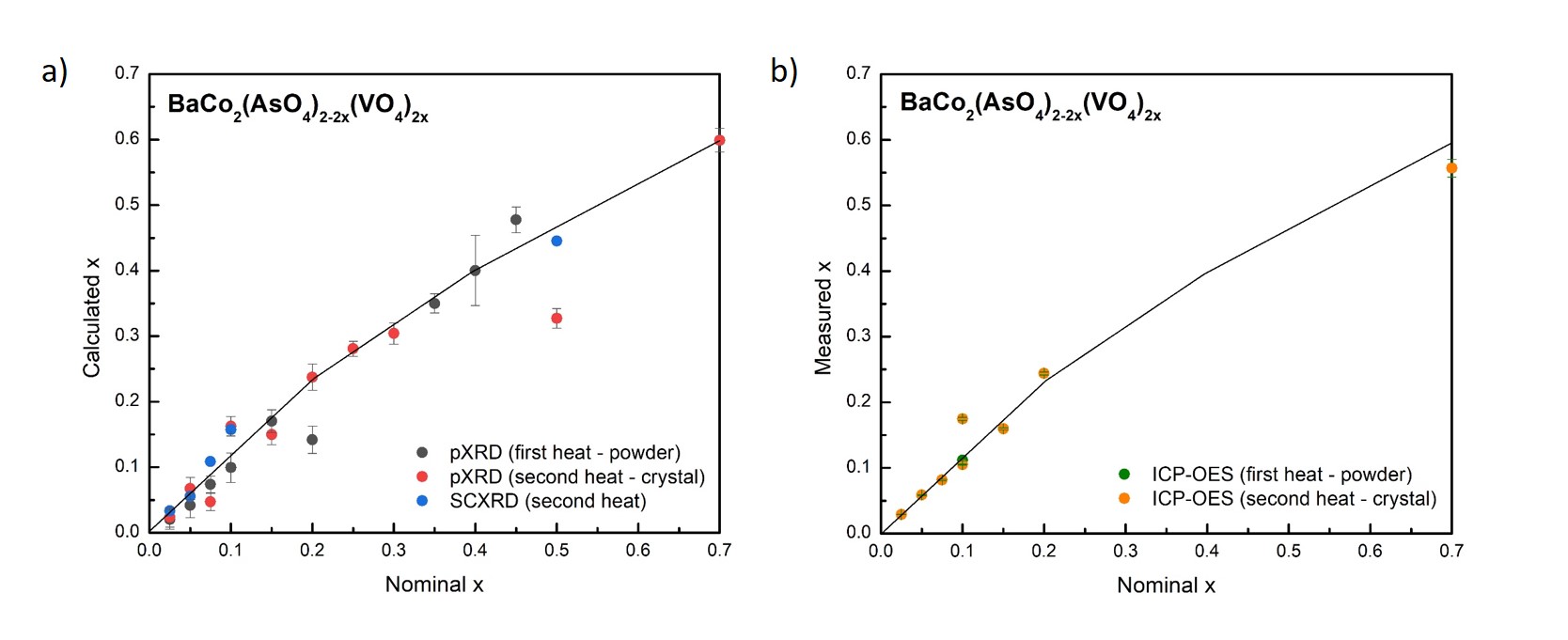}
  \caption{\textbf{Calculated and measured V-content of representative BCAO-V compositions.} a) Calculated V-content of several BCAO-V compositions, derived from Rietveld refinement of once- (black) and twice-heated (red) sample pXRD patterns. V-content was also estimated from SCXRD refinement data for twice-heated BCAO-V crystals (blue) using: $x = (33-33z)/10$, where $x$ is the estimated V-content for each sample and $z$ is the refined occupancy factor for As. This accounts for the 10-electron difference between arsenic and vanadium, allowing for a more reliable estimate of V-occupancy on the As-site. b) V-content of once (green) and twice-heated (orange) samples measured via ICP-OES analysis. The measured V-content is in good agreement with that estimated via refinement of pXRD and SCXRD data, as shown by the black trend line, and suggests an upper substitutional limit of $\sim$56\%.}
  \label{V-incorporation}
\end{figure}
\newpage 

\section{Magnetic characterization}

\begin{figure}
  \includegraphics[width=1.00\textwidth]{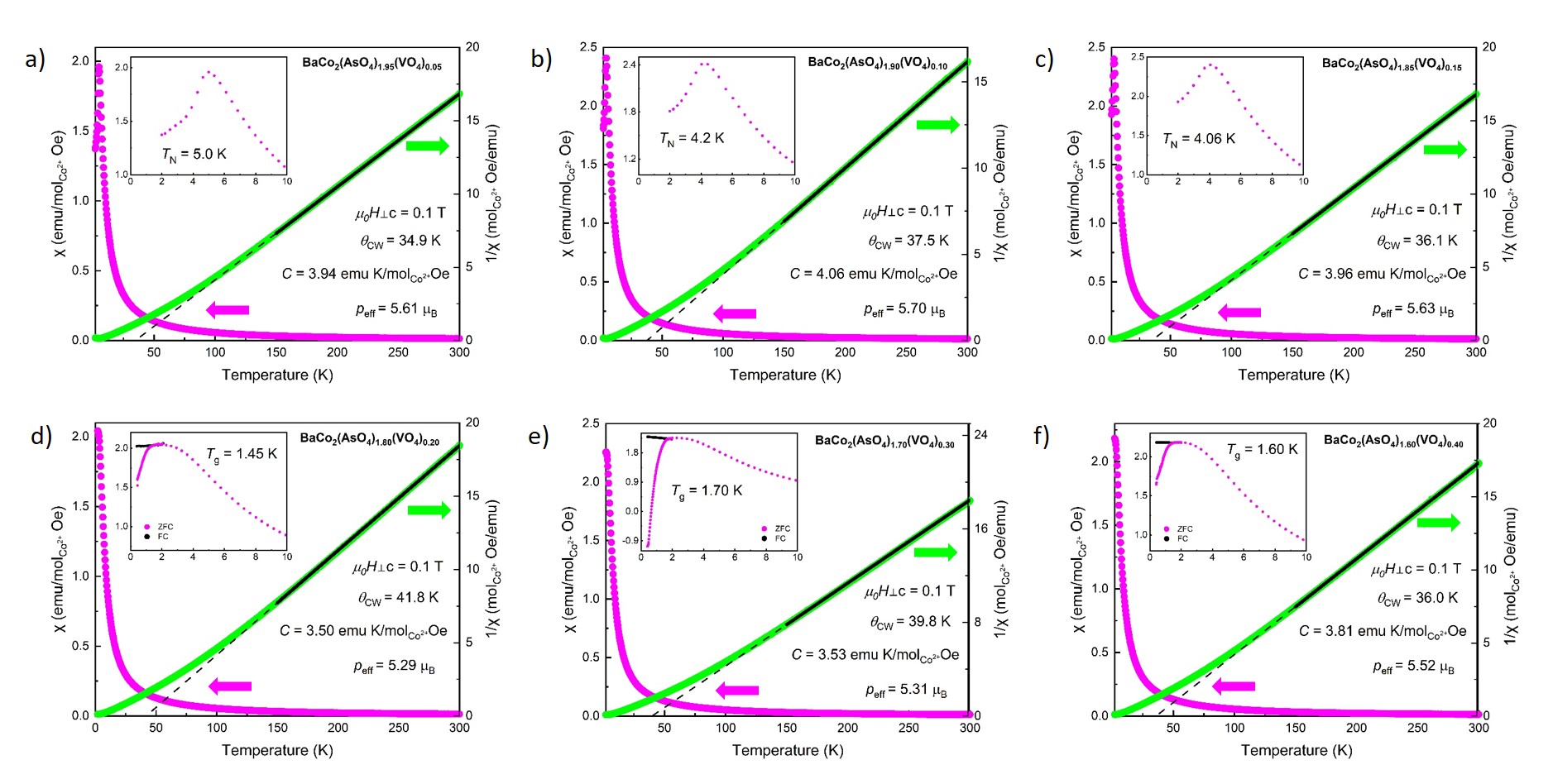}
  \caption{\textbf{In-plane, temperature-dependent magnetic (pink) and inverse magnetic (green) susceptibility of representative a) x~=~0.025, b) x~=~0.05, c) x~=~0.075, d) x~=~0.10, e) x~=~0.15, and f) x~=~0.20 BCAO-V single crystals.} Best fit lines are shown in black and are extrapolated to \emph{T}~=~0~K. Curie Weiss analysis from \emph{T}~=~150-300~K suggests dominant ferromagnetic interactions between \ch{Co^{2+}} ions within the honeycomb plane. Calculated Curie-Weiss parameters for each composition are also summarized in Table 1. Only minor variations are observed in both the calculated $\theta_{CW}$ and $p_{eff}$, suggesting that up to 20\% nominal substitution of the apical arsenate group has little impact on the strength of the in-plane superexchange correlations. Zero-field-cooled (ZFC - pink) and field-cooled (FC - black) magnetic susceptibility of the x~=~0.10-0.20 compositions from \emph{T} = 0.4-2.0 K (insets of d-f). Relative to the shifts observed in the calculated out-of-plane CW parameters, less varied behavior is observed when the field is applied within the honeycomb plane, suggesting that increasing V-substitution has a minimal impact on the relative strength of the in-plane superexchange interactions.}
  \label{MvT_in_plane}
\end{figure}
\newpage


\begin{figure}
  \includegraphics[width=1.0\textwidth]{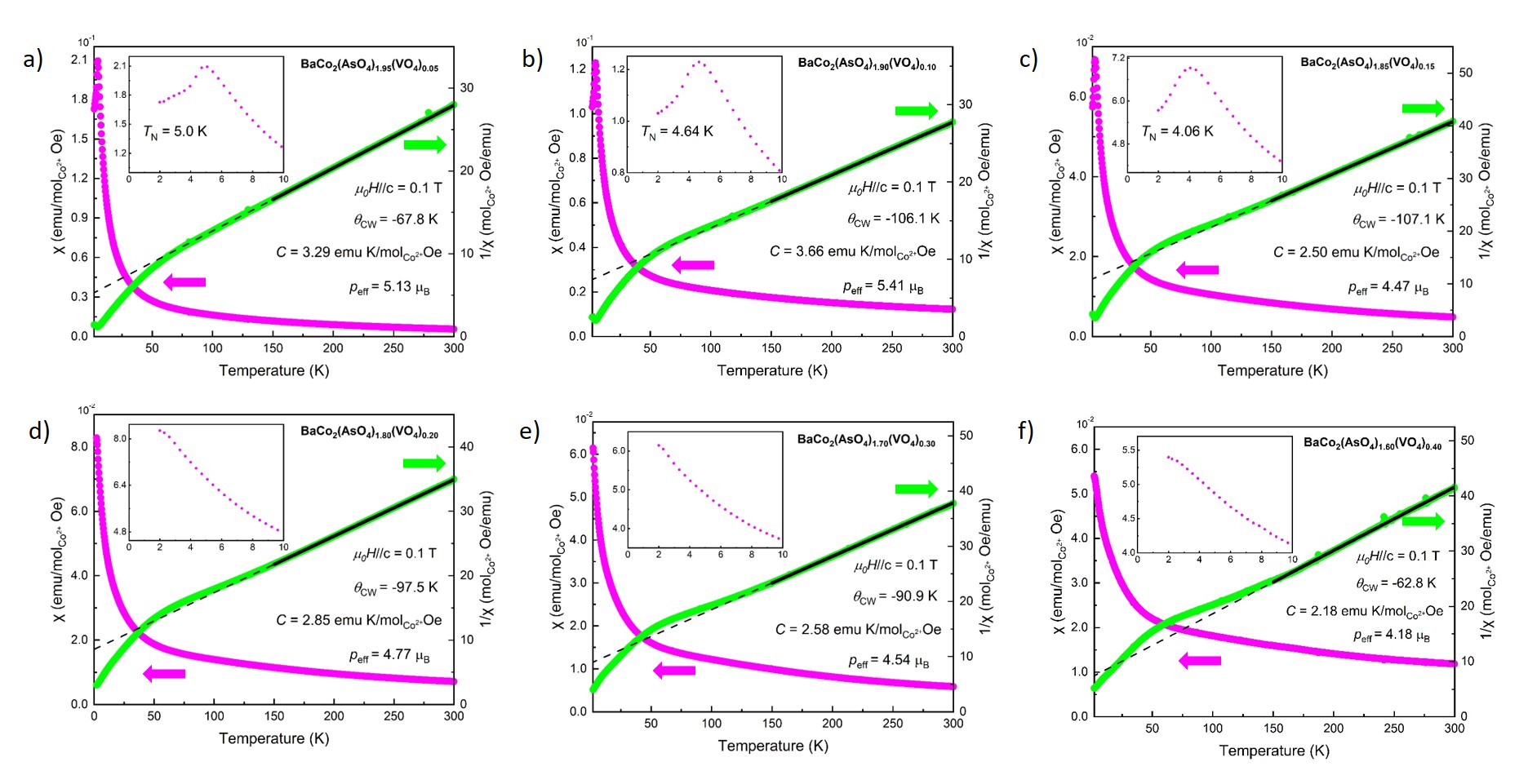}
  \caption{\textbf{Out-of-plane, temperature-dependent magnetic (pink) and inverse magnetic (green) susceptibility of representative a) x~=~0.025, b) x~=~0.05, c) x~=~0.075, d) x~=~0.10, e) x~=~0.15, and f) x~=~0.20 BCAO-V single crystals.} Curie Weiss analysis from \emph{T}~=~150-300~K suggests dominant antiferromagnetic interactions between \ch{Co^{2+}} ions in adjacent layers. Best fit lines are shown in black and are extrapolated to \emph{T}~=~0~K. Calculated Curie-Weiss parameters for each composition are also summarized in Table S12. Relative to the shifts observed in the calculated in-plane CW parameters, more varied behavior is observed when the field is applied along the stacking axis, suggesting that increasing V-substitution has a larger impact on the longer, out-of-plane magnetic exchange pathways.}
  \label{MvT_out_of_plane}
\end{figure}
\newpage

\begin{table}
    \centering
    \caption{\textbf{Fitting parameters obtained from analysis of BCAO-V single crystal magnetization measurements ($\mu_0\text{H}\parallel$ c) from \emph{T}~=~150-300~K.}}
    \label{MvT_para_CW}
    \begin{tabular}{c|c|c|c|c|c}
    \hline
    x & DC \emph{T}$_\text{N}$ & \emph{T}$_\text{f}$ & $\theta_{\text{CW}}$ & $p_{\text{eff}}$ & $\chi_0$ \\ 
     & (K) & (K) & (K) & ($\mu_\text{B}$) & \\\hline
    0 & 5.40 & - & -167.7 & 5.91 & - \\
    0.025 & 5.00 & 3.67 & -67.8(1.3) & 5.13(2) & -0.007 \\
    0.05 & 4.64 & 3.23 & -106.1(8) & 5.41(1) & 0.007 \\
    0.075 & 4.06 & 2.98 & -107.1(1.3) & 4.47(2) & -0.003 \\
    0.10 & - & - & -97.5(1) & 4.47(1) & 0 \\
    0.15 & - & - & -90.9(1.1) & 4.54(1) & -0.001 \\
    0.20 & - & - & -62.8(2.8) & 4.17(1) & 0.012 \\ \hline
    \end{tabular}
\end{table}
\newpage

\begin{figure}
  \includegraphics[width=0.44\textwidth]
  {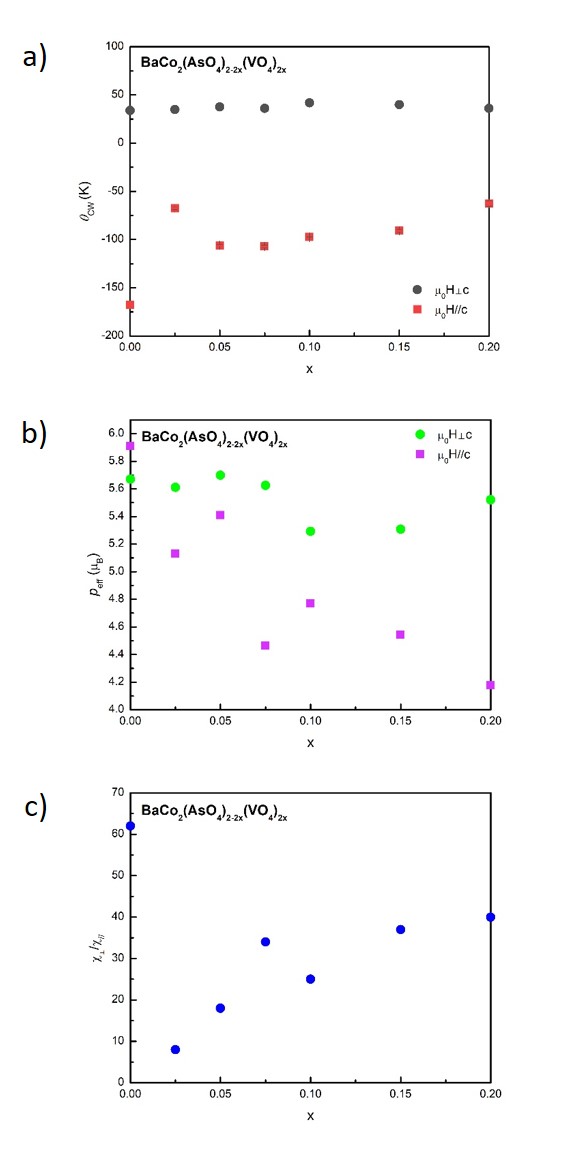}
\caption{\textbf{Calculated a) Curie-Weiss temperature $\theta_{CW}$ and b) effective magnetic moment p$_{eff}$ of BCAO-V single crystals as a function of nominal V-substitution with $\mu_0H\perp$ c (black, green circles) and $\mu_0H\parallel$ c (red, purple squares).} When the field is applied within the honeycomb plane, minimal variation is observed in both $\theta_{CW}$ and p$_{eff}$ with increasing V-content. When the field is instead applied along the stacking axis, more varied behavior is observed. c) The anisotropic factor, $\chi_{\perp}/\chi_{\parallel}$, as a function of nominal V-substitution. The anisotropic factor is defined as the magnitude of the measured magnetic susceptibility with the field applied perpendicular to the stacking axis divided by that measured with the field applied parallel to the same axis.}
  \label{CW_params}
\end{figure}
\newpage


\begin{figure}
  \includegraphics[width=1.00\textwidth]{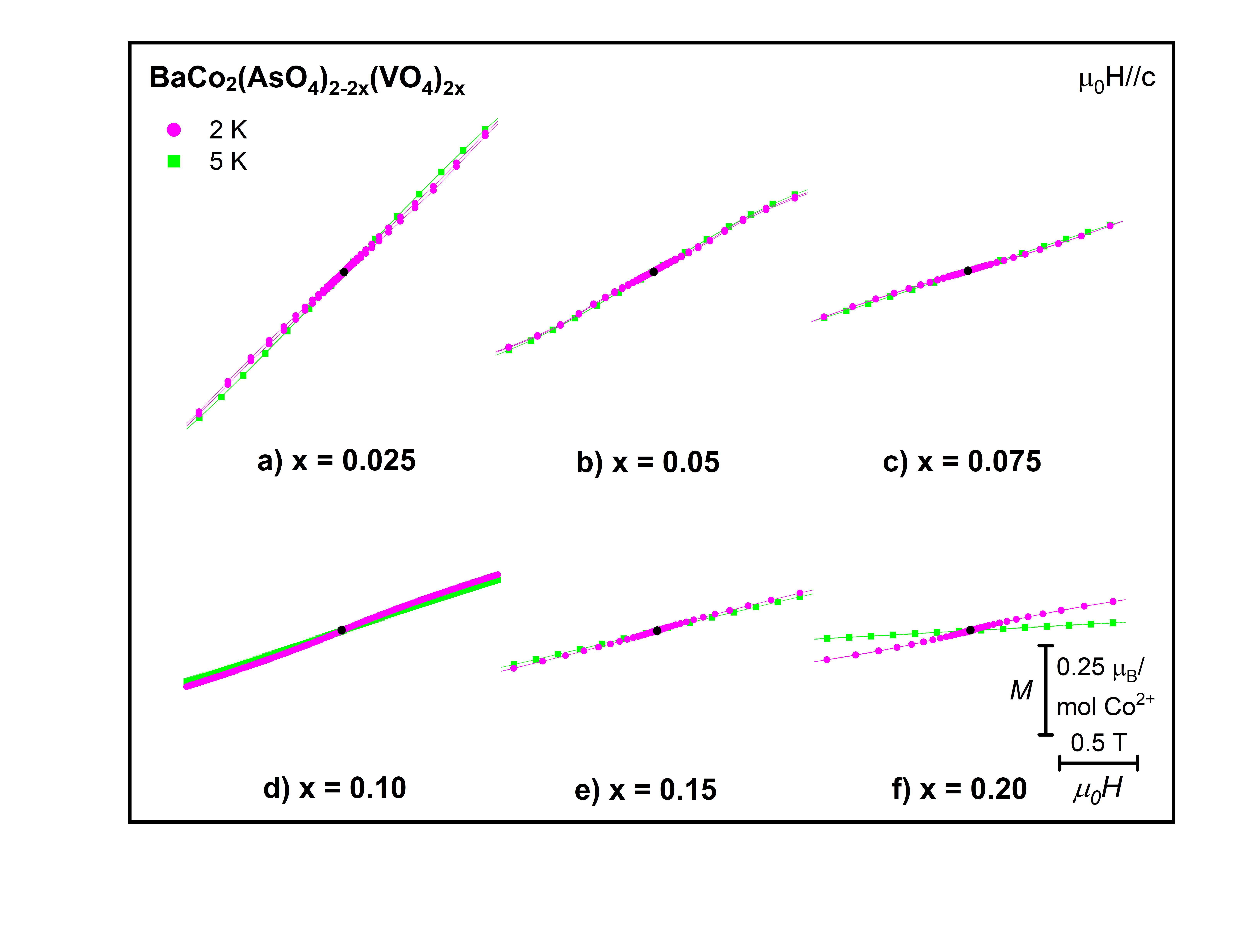}
  \caption{\textbf{Out-of-plane, field-dependent magnetization of representative a) x~=~0.025, b) x~=~0.05, c) x~=~0.075, d) x~=~0.10, e) x~=~0.15, and f) x~=~0.20 BCAO-V single crystals at \emph{T}~=~2.0~K (pink) and \emph{T}~=~5.0~K (green).} The magnitude of the magnetization decreases with increasing V-content and no coercivity is observed for any composition.}
  \label{MvH_out_of_plane}
\end{figure}
\newpage


\begin{figure}
  \includegraphics[width=1.00\textwidth]{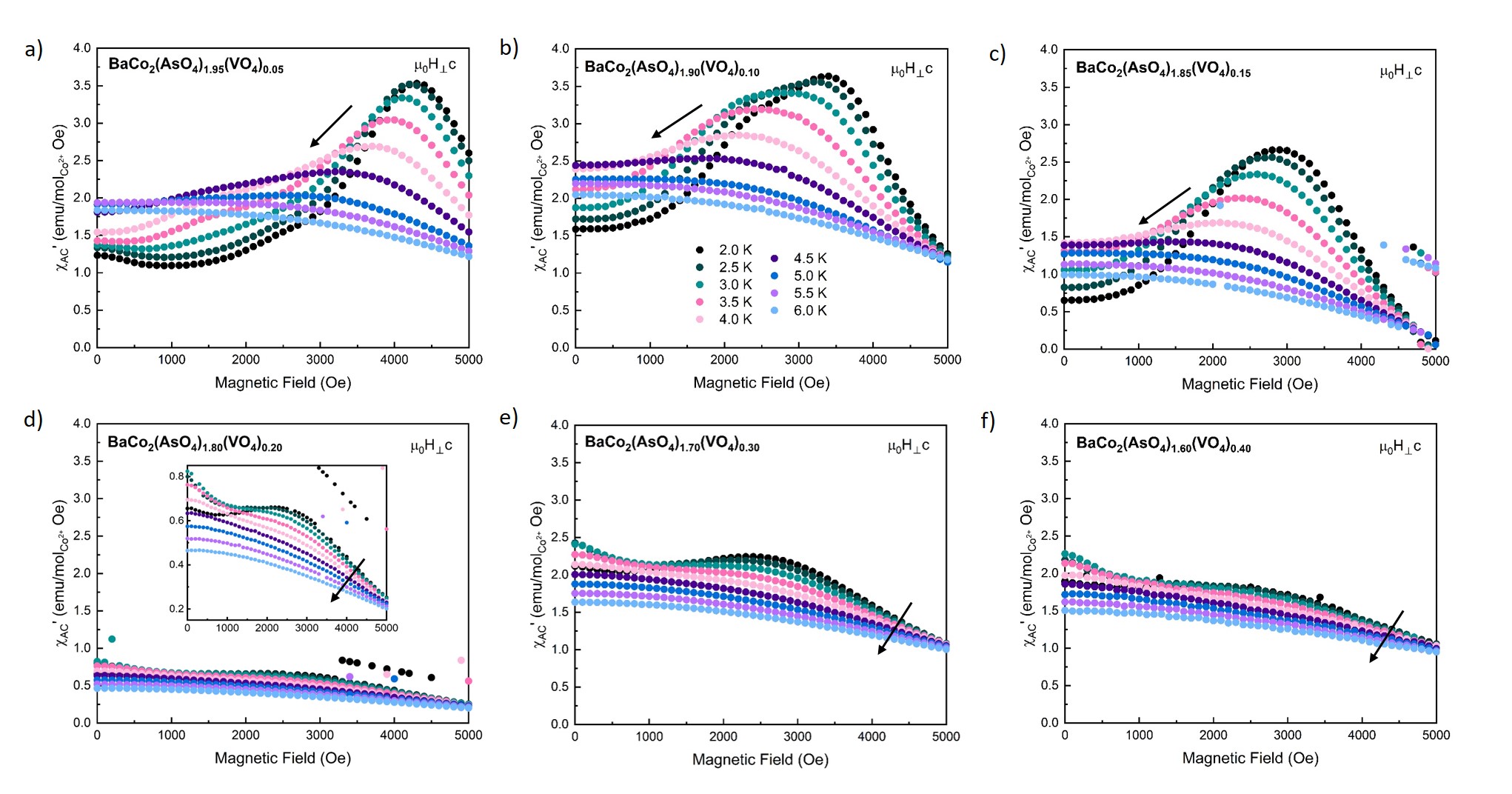}
  \caption{\textbf{Real component of the AC magnetic susceptibility of a) x~=~0.025, b) x~=~0.05, c) x~=~0.075, d) x~=~0.10, e) x~=~0.15, and f) x~=~0.20 BCAO-V single crystal compositions as a function of applied in-plane magnetic field from \emph{T}~=~2.0-6.0~K.} All measurements were performed with an applied AC field of frequency $\omega$ = 775 Hz. With increasing V-content, the upper critical field, H$_{c1}$, observed in unsubstituted BCAO gradually decreases to meet the lower critical field H$_{c2}$, which is largely unchanged from the parent material. Above x~=~0.075, only the transition at H$_{c2}$ is visible. This is consistent with a gradual tuning of the incommensurate ground state, leaving only the commensurate ground state at higher substitution levels. The magnitude of the AC susceptibility also steadily decreases across the series, with a larger drop for the x~=~0.10 composition (inset provided for clarity), however the general behavior remains consistent.}
  \label{AC_chi_vs_B}
\end{figure}
\newpage


\begin{figure}
  \includegraphics[width=1.0\textwidth]
  {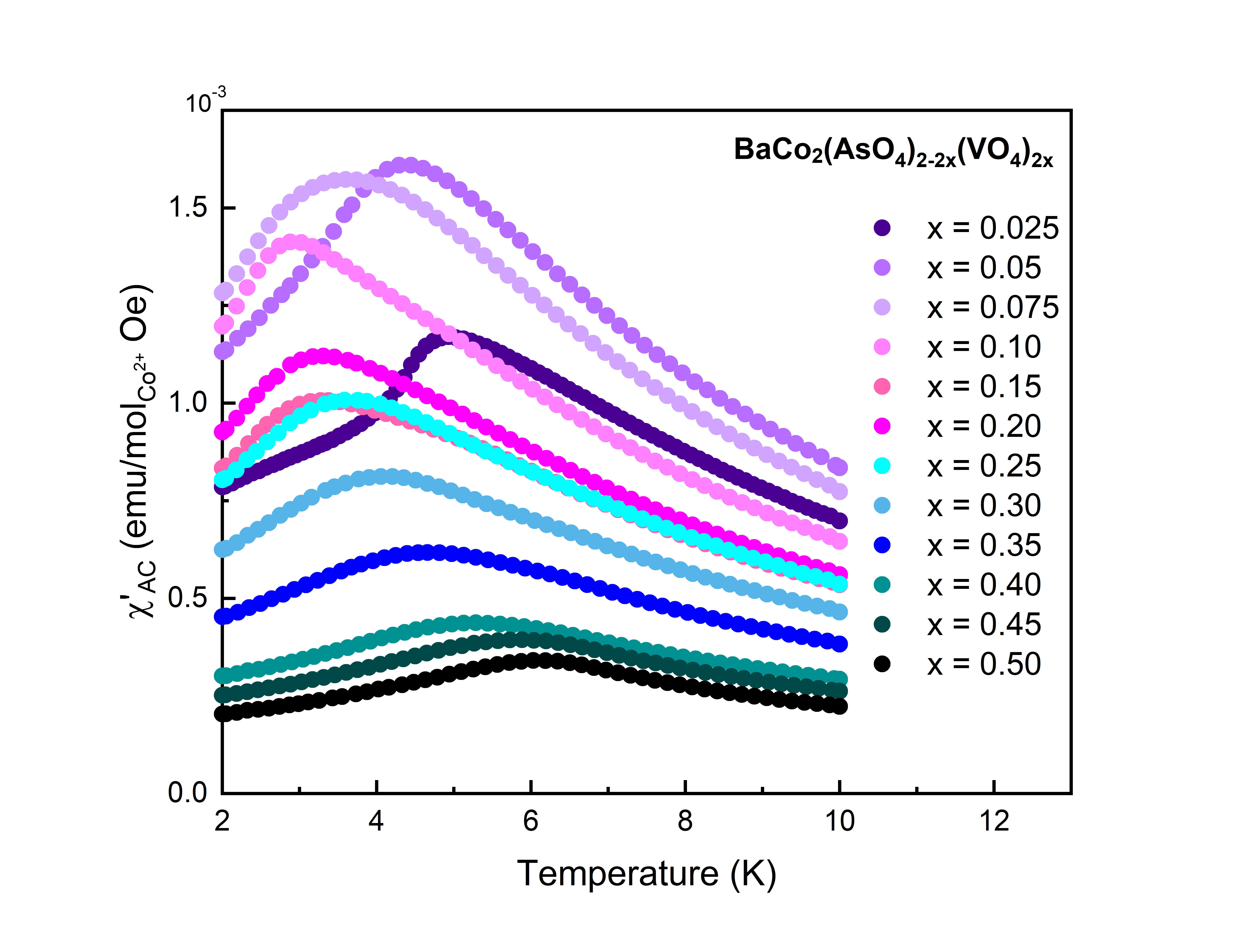}
  \caption{\textbf{Real component of the AC magnetic susceptibility of BCAO-V powder compositions as a function of V-substitution from \emph{T}~=~2.0-10~K.} Up to x~=~0.10, the observed transition is suppressed to about \emph{T}~=~2.9~K, before gradually broadening and increasing in temperature with further vanadium incorporation. All measurements were performed with an applied AC field of frequency $f$ = 600 Hz.}
  \label{AC_chi'}
\end{figure}
\newpage


\begin{figure}
  \includegraphics[width=1.0\textwidth]{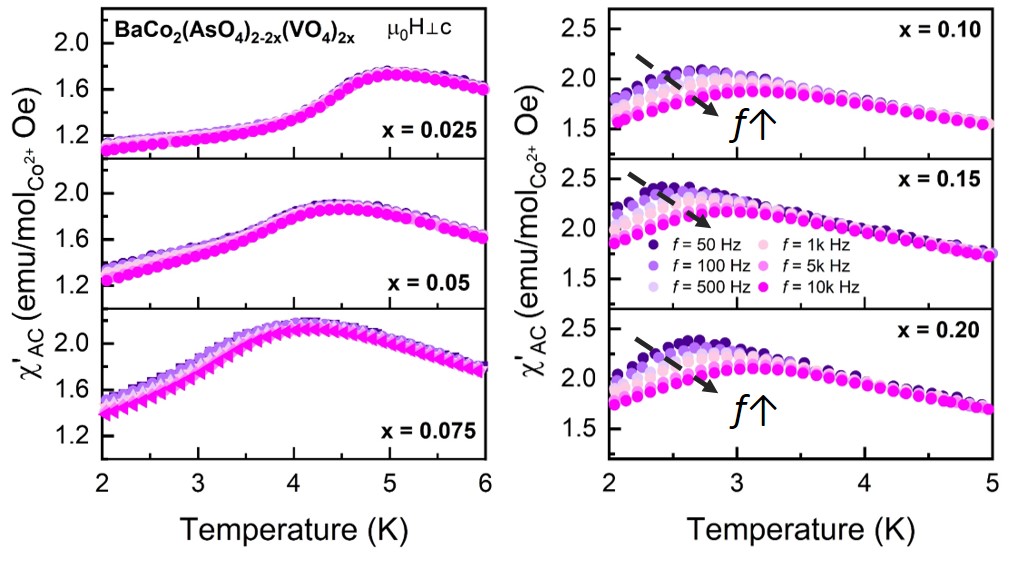}
  \caption{\textbf{Zoomed view of the real component of the AC magnetic susceptibility of (left, from top to bottom) x~=~0.025, x~=~0.05, x~=~0.075, and (right, from top to bottom) x~=~0.10, x~=~0.15, x~=~0.20 BCAO-V single crystals as a function of temperature and with an applied AC field of frequency $f$ = 50-10000 Hz (purple to pink).} From x~=~0.025-0.075 nominal V-substitution, the transition temperature is observed to gradually decrease, but is frequency-independent. For these compositions, an increasing frequency dependence is observed only below the transition temperature. Above this level of substitution, the magnitude of the measured susceptibility briefly decreases and an increasing frequency dependence of the transition temperature is observed. For the x~=~0.10 sample, a variable frequency-dependence of the AC susceptibility is observed across different batches, supporting the proximity of this composition to a critical point. The x~=~0.15 and x~=~0.20 samples typically display a greater frequency dependence relative to the x~=~0.10 composition, up to $f$~=~10000~Hz. The region of these frequency shifts are denoted by the dashed black lines.} 
  \label{AC_chi_vs_T}
\end{figure}
\newpage

\begin{figure}
  \includegraphics[width=1.00\textwidth]{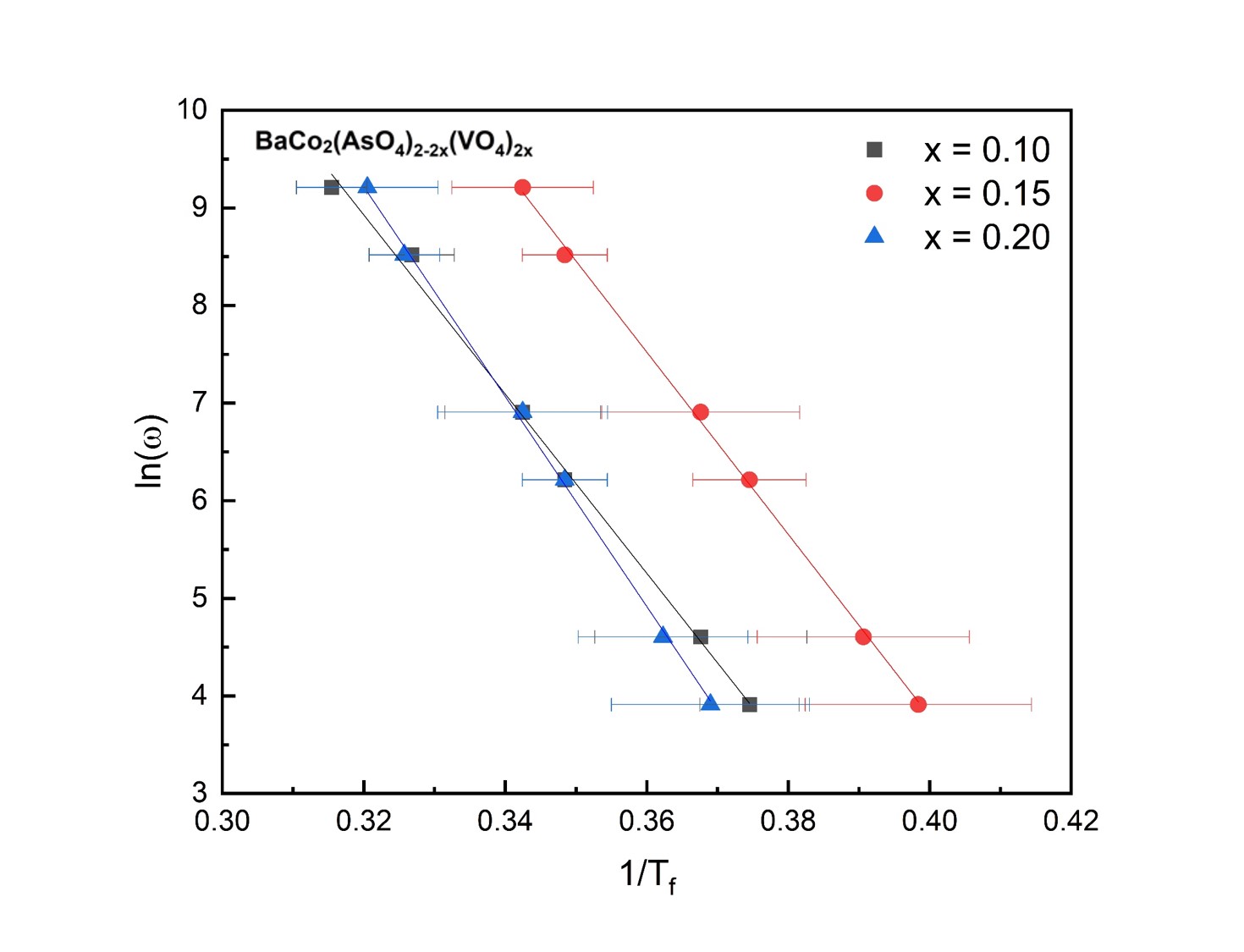}
  \caption{\textbf{Plot of 1/T vs. ln($\omega$) of x~=~0.10 (black), x~=~0.15 (red), and x~=~0.20 (blue) BCAO-V single crystals, derived from in-plane AC magnetic susceptibility measurements.} Across the full frequency range of $f$~=~50-10000 Hz, all three compositions exhibit a linear dependence suggestive of a cluster glass magnetic ground state. The more variable frequency dependence of the x~=~0.10 composition observed across batches indicates a more complex ground state spin configuration and further supports its proximity to a compositional critical point in the series.}
  \label{AC_freq_dep}
\end{figure}
\newpage

\begin{table}
    \centering
    \begin{tabular}{c|c|c|c}
    \hline
    x & \emph{T}$_\text{f}$ ($f$~=~50~Hz) (K) & E$_\text{a}$/k$_B$ (K) & $\omega_0$ (Hz) \\ \hline
    0.10 & 2.67 $\pm$ 0.007 K & 91.8 $\pm$ 23.8 K & 1.21 x 10$^{13}$-1.54 x 10$^{20}$\\
    0.15 & 2.51 $\pm$ 0.016 K & 93.3 $\pm$ 29.9 K & 1.27 x 10$^{13}$-4.11 x 10$^{22}$ \\
    0.20 & 2.71 $\pm$ 0.014 K & 107.5 $\pm$ 34.6 K & 1.27 x 10$^{13}$-1.21 x 10$^{24}$ \\ \hline
    \end{tabular}
    \caption{\textbf{Arrhenius-type fitting parameters for x~=~0.10, x~=~0.15, and x~=~0.20 BCAO-V single crystals, derived from in-plane ($\mu_0\text{H}\perp$ c) AC magnetic susceptibility measurements with associated statistical errors}.}
    \label{Arrhenius}
\end{table}
\newpage

\section{Heat capacity data}

\begin{figure}
  \includegraphics[width=1.0\textwidth]{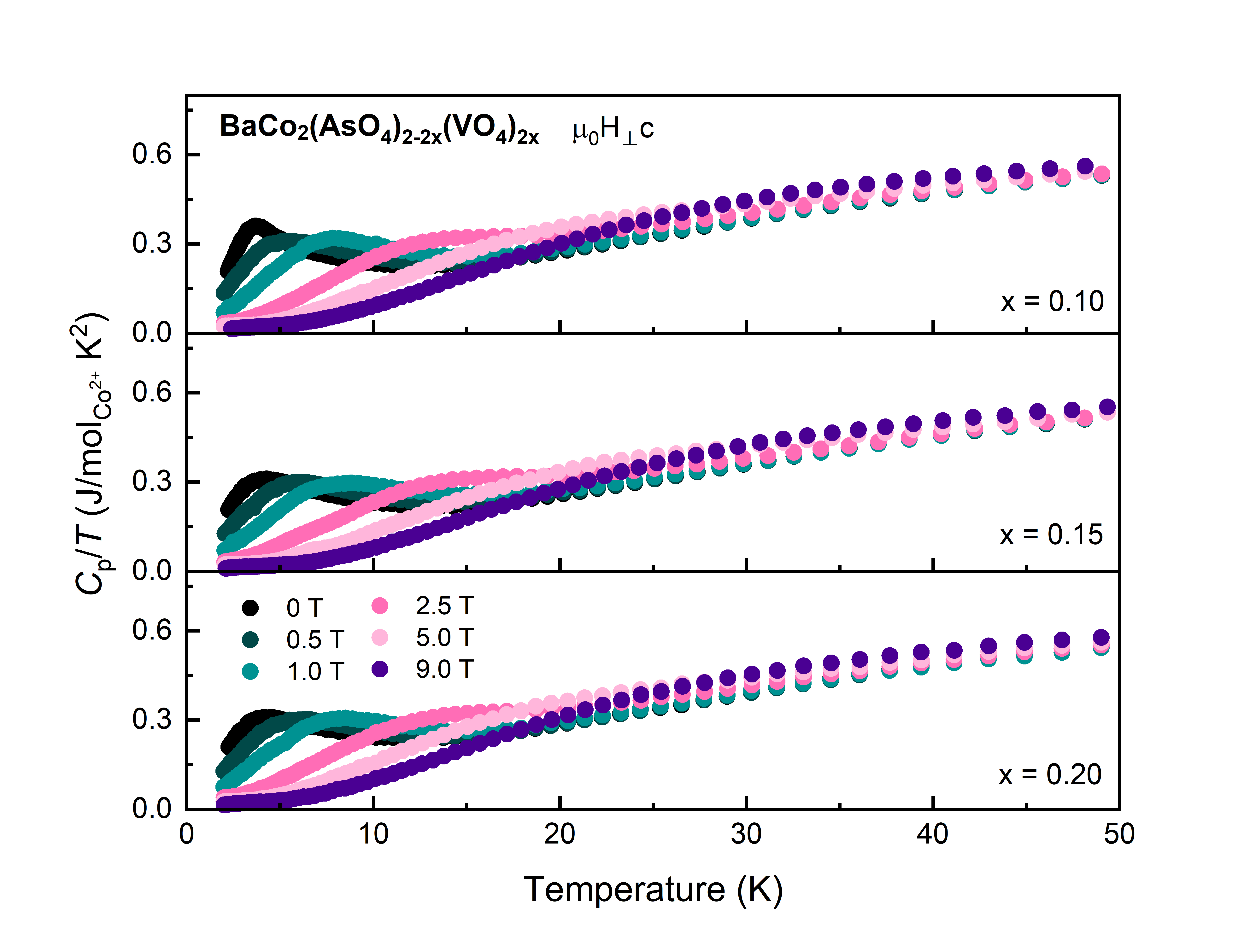}
  \caption{\textbf{Representative heat capacity of (from top to bottom) x~=~0.10, x~=~0.15, and x~=~0.20 BCAO-V single crystals as a function of temperature in a $\mu_0\text{H}$~=~0-9~T magnetic field applied within the honeycomb plane.} As opposed to the more varied behavior observed when the field is applied along the stacking axis, when it is instead applied within the \emph{ab} plane, similar suppression is observed for all three compositions with increasing applied field.}
  \label{HC_SC_perp}
\end{figure} \newpage


\begin{figure}
  \includegraphics[width=1.0\textwidth]{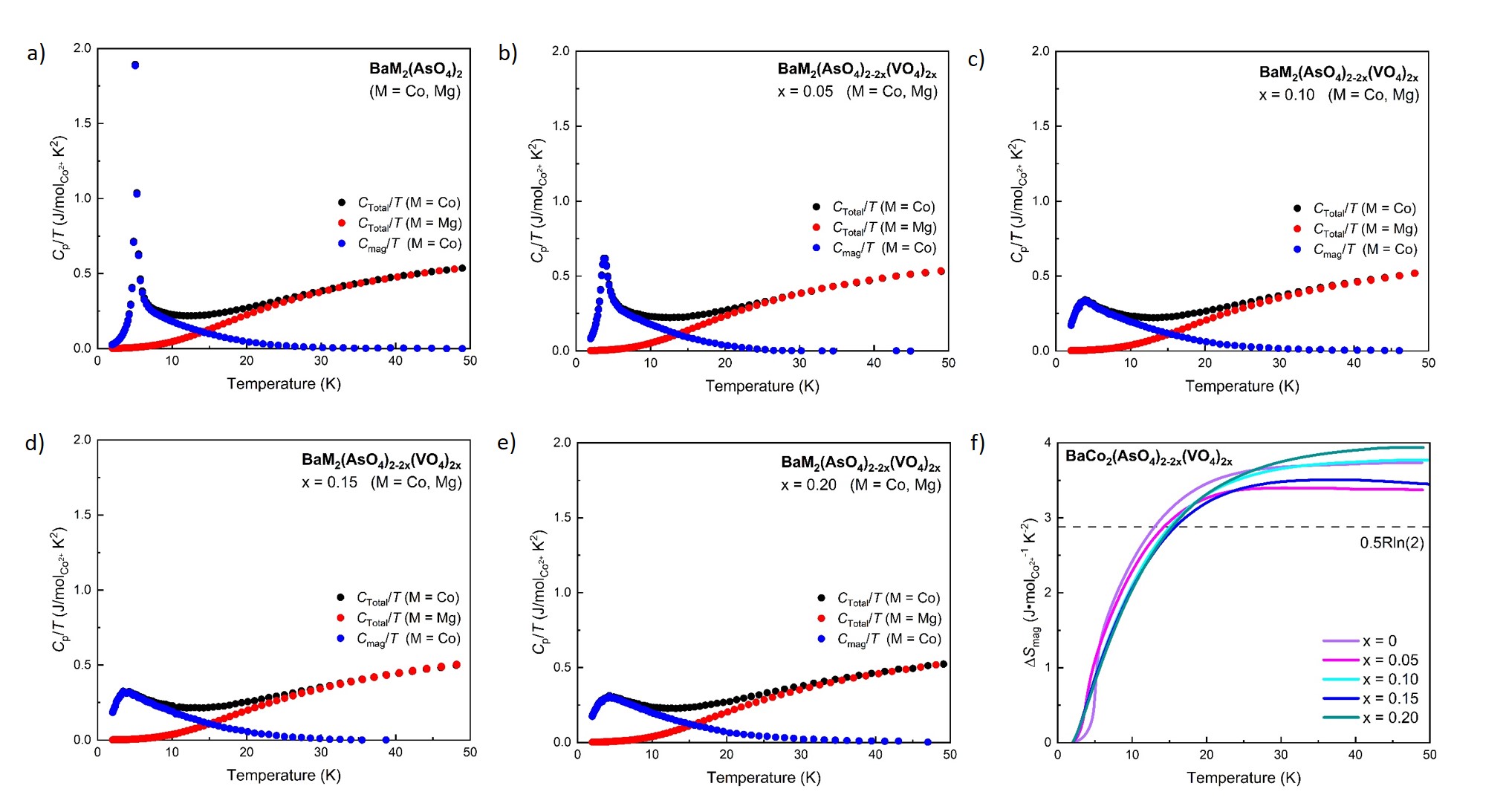}
  \caption{\textbf{Zero-field heat capacity divided by temperature as a function of temperature of representative a) x~=~0, b) x~=~0.05, c) x~=~0.10, d) x~=~0.15, and e) x~=~0.20 single crystal BCAO-V (black) and non-magnetic BMAO-V (red) compositions.} As the measured vibrational frequencies of the two phases differ considerably, the non-magnetic analogue could not be used to reliably approximate the phonon contribution to the measured heat capacity. However, in order to obtain a rough estimate of the magnetic contribution (blue), the Mg-analogue was scaled by a factor of a) 1.20, b) 1.29, c) 1.44, d) 1.38, and e) 1.54 and subtracted from that of the Co-analogue by \emph{T}~=~50~K. Integration of this estimated contribution over the same temperature range yields an entropy rise greater than the predicted 1/2Rln2 for an ideal Kitaev system and less than the Rln2 expected for a typical S~=~1/2 magnet.}
  \label{Mag_non-mag_HC}
\end{figure}